\documentclass[twocolumn,longauth]{aa} 

\usepackage{ulem} 
\usepackage{natbib,twoopt}
\usepackage[varg]{txfonts} 
\bibpunct{(}{)}{;}{a}{}{,}
\usepackage{booktabs} 

\usepackage{graphicx}
\usepackage{float}
\usepackage{lscape}
\usepackage{pdflscape}
\usepackage{sidecap}
\usepackage{xcolor}
\usepackage{comment}
\usepackage{rotating}
\usepackage{siunitx}
\usepackage{soul}
\usepackage{cancel}
\usepackage{microtype}
\usepackage{tabularx}
\usepackage{longtable}
\usepackage{array}
\usepackage{makecell}

\usepackage{caption}

\usepackage{amssymb}
\usepackage[utf8]{inputenc}

\usepackage{academicons}
\definecolor{orcidlogocol}{HTML}{A6CE39}
\usepackage{placeins}
\usepackage{subfigure} 
\usepackage{hyperref}

\AtBeginDocument{%
  \abovedisplayskip=8pt plus 2pt minus 4pt
  \belowdisplayskip=8pt plus 2pt minus 4pt
  \abovedisplayshortskip=4pt plus 2pt
  \belowdisplayshortskip=4pt plus 2pt minus 2pt
}

\defcitealias{alerts_classifier}{SS21a}
\defcitealias{sanchez_ZTF_2023}{SS23}
\defcitealias{arevalo2025}{A26}
\defcitealias{VARPZ}{SS26}

\begin{document}
\title{\texttt{VAR-PZnn}: A machine-learning framework for AGN photometric redshifts using color and variability-based features}
\titlerunning{A machine-learning framework for AGN photometric redshifts using color and variability-based features}
\author{
S. Satheesh-Sheeba \inst{1,3}\thanks{E-mail: 1998sarath.ss@gmail.com}~\href{https://orcid.org/0009-0003-0654-6805}{\includegraphics[scale=0.05]{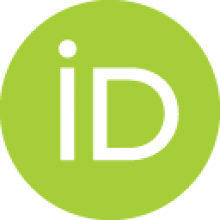}} \and
P. Sánchez-Sáez\inst{2} ~\href{https://orcid.org/0000-0003-0820-4692}{\includegraphics[scale=0.05]{Figures/orcid-ID.png}}\and
R. J. Assef\inst{3}~\href{https://orcid.org/0000-0002-9508-3667}{\includegraphics[scale=0.05]{Figures/orcid-ID.png}}\and
T. Anguita\inst{1}~\href{https://orcid.org/0000-0003-0930-5815}{\includegraphics[scale=0.05]{Figures/orcid-ID.png}} \and
R. Shirley \inst{4}~\href{https://orcid.org/0000-0002-1114-0135}{\includegraphics[scale=0.05]{Figures/orcid-ID.png}} \and
M. Salvato \inst{4}~\href{https://orcid.org/0000-0001-7116-9303}{\includegraphics[scale=0.05]{Figures/orcid-ID.png}} \and 
P. Arévalo \inst{5,6}~\href{https://orcid.org/0000-0001-5675-6323}{\includegraphics[scale=0.05]{Figures/orcid-ID.png}} \and
T T. Ananna \inst{7} ~\href{https://orcid.org/0000-0001-8211-3807}{\includegraphics[scale=0.05]{Figures/orcid-ID.png}} \and
F. E. Bauer \inst{8}~\href{https://orcid.org/0000-0002-8686-8737}{\includegraphics[scale=0.05]{Figures/orcid-ID.png}} \and
C. G. Bornancini \inst{9,10}~\href{https://orcid.org/0000-0001-6800-3329}{\includegraphics[scale=0.05]{Figures/orcid-ID.png}}\and
W.N. Brandt \inst{11,12,13} ~\href{https://orcid.org/0000-0002-0167-2453}{\includegraphics[scale=0.05]{Figures/orcid-ID.png}} \and
D. De Cicco \inst{14,15}~\href{https://orcid.org/0000-0001-7208-5101}{\includegraphics[scale=0.05]{Figures/orcid-ID.png}} \and
M. Espinoza-Ortiz \inst{1}~\href{https://orcid.org/0009-0002-6369-6255}{\includegraphics[scale=0.05]{Figures/orcid-ID.png}} \and
J. Fagin \inst{16,17,18}~\href{https://orcid.org/0000-0001-8723-6136}{\includegraphics[scale=0.05]{Figures/orcid-ID.png}} \and
M. Fatovi\'c\inst{15,14,19}~\href{https://orcid.org/0000-0003-1911-4326}{\includegraphics[scale=0.05]{Figures/orcid-ID.png}} \and
A. W. Graham \inst{20}~\href{https://orcid.org/0000-0002-6496-9414}{\includegraphics[scale=0.05]{Figures/orcid-ID.png}} \and
H. Guo  \inst{21}~\href{https://orcid.org/0000-0001-8416-7059}{\includegraphics[scale=0.05]{Figures/orcid-ID.png}}\and
L. Hernandez-García \inst{3,22}~\href{https://orcid.org/0000-0002-8606-6961}{\includegraphics[scale=0.05]{Figures/orcid-ID.png}}\and 
D. Ili\'c\inst{23,24}~\href{https://orcid.org/0000-0002-1134-4015}{\includegraphics[scale=0.05]{Figures/orcid-ID.png}} \and
A. B. Kova\v cevi\'c \inst{23}~\href{https://orcid.org/0000-0001-5139-1978}{\includegraphics[scale=0.05]{Figures/orcid-ID.png}}\and
P. Lira \inst{25,6}~\href{https://orcid.org/0000-0003-1523-9164}{\includegraphics[scale=0.05]{Figures/orcid-ID.png}}\and
A. I. Malz \inst{26}~\href{https://orcid.org/0000-0002-8676-1622}{\includegraphics[scale=0.05]{Figures/orcid-ID.png}}\and 
M. Marculewicz \inst{7} ~\href{https://orcid.org/0000-0002-1380-1785}{\includegraphics[scale=0.05]{Figures/orcid-ID.png}} \and
D. Marsango\inst{27,28}~\href{https://orcid.org/0000-0001-5465-0824}{\includegraphics[scale=0.05]{Figures/orcid-ID.png}} \and
C. Mazzucchelli\inst{3}~\href{https://orcid.org/0000-0002-5941-5214}{\includegraphics[scale=0.05]{Figures/orcid-ID.png}} \and
T. Mkrtchyan \inst{3} ~\href{https://orcid.org/0009-0000-6071-4353}{\includegraphics[scale=0.05]{Figures/orcid-ID.png}} \and
S. Panda \inst{29}~\href{https://orcid.org/0000-0002-5854-7426}{\includegraphics[scale=0.05]{Figures/orcid-ID.png}}\and
A. Peca \inst{3}~\href{https://orcid.org/0000-0003-2196-3298}{\includegraphics[scale=0.05]{Figures/orcid-ID.png}}  \and
V. Petrecca \inst{15}~\href{https://orcid.org/0000-0002-3078-856X}{\includegraphics[scale=0.05]{Figures/orcid-ID.png}}  \and 
B. Rani \inst{30,31}~\href{https://orcid.org/0000-0001-5711-084X}{\includegraphics[scale=0.05]{Figures/orcid-ID.png}}\and
C. Ricci \inst{32}~\href{https://orcid.org/0000-0001-5231-2645}{\includegraphics[scale=0.05]{Figures/orcid-ID.png}}\and
G. T. Richards \inst{33}~\href{https://orcid.org/0000-0002-1061-1804}{\includegraphics[scale=0.05]{Figures/orcid-ID.png}} \and
R. A. Riffel \inst{27,28}~\href{https://orcid.org/0000-0003-0483-3723}{\includegraphics[scale=0.05]{Figures/orcid-ID.png}}\and
A. Rojas-Lilayú \inst{34}~\href{https://orcid.org/0000-0003-0006-8681}{\includegraphics[scale=0.05]{Figures/orcid-ID.png}}  \and
E. Saremi \inst{35}~\href{https://orcid.org/0000-0002-5075-1764}{\includegraphics[scale=0.05]{Figures/orcid-ID.png}}\and
D. P. Schneider \inst{11,12}~\href{https://orcid.org/0000-0001-7240-7449}{\includegraphics[scale=0.05]{Figures/orcid-ID.png}} \and
B. Sotomayor\inst{5} ~\href{https://orcid.org/0009-0001-1495-6068}{\includegraphics[scale=0.05]{Figures/orcid-ID.png}}  \and 
M. J. Temple \inst{36} ~\href{https://orcid.org/0000-0001-8433-550X}{\includegraphics[scale=0.05]{Figures/orcid-ID.png}} \and
A. Viitanen \inst{32,15} ~\href{https://orcid.org/0000-0001-9383-786X}{\includegraphics[scale=0.05]{Figures/orcid-ID.png}}\and
I. Yoon \inst{37,38} ~\href{https://orcid.org/0000-0001-9163-0064}{\includegraphics[scale=0.05]{Figures/orcid-ID.png}}\and
Z. Yu \inst{11,12} ~\href{https://orcid.org/0000-0002-6990-9058}{\includegraphics[scale=0.05]{Figures/orcid-ID.png}}\and 
F. Zou \inst{39}~\href{https://orcid.org/0000-0002-4436-6923}{\includegraphics[scale=0.05]{Figures/orcid-ID.png}}
}

\institute{Instituto de Astrof\'{i}sica, Facultad de Ciencias Exactas, Universidad Andr\'{e}s Bello, Fern\'{a}ndez Concha 700, 7591538, Las Condes, Santiago, Chile
\and
European Southern Observatory, Karl-Schwarzschild-Strasse 2, 85748 Garching bei M\"unchen, Germany
\and
Instituto de Estudios Astrof\'{i}sicos, Facultad de Ingenier\'{i}a y Ciencias, Universidad Diego Portales, Av. Ej\'{e}rcito Libertador 441, Santiago, Chile
\and
Max-Planck-Institut f\"ur extraterrestrische Physik, Giessenbachstr. 1, 85748 Garching, Germany
\and
Instituto de F\'{i}sica y Astronom\'{i}a, Facultad de Ciencias, Universidad de Valpara\'{i}so, Gran Breta\~{n}a 1111, Valpara\'{i}so, Chile
\and
Millennium Nucleus on Transversal Research and Technology to Explore Supermassive Black Holes (TITANS), Valparaiso, Chile
\and
Department of Physics and Astronomy, Wayne State University, 666 W. Hancock St, Detroit, MI, 48201, USA
\and
Instituto de Alta Investigaci\'{o}n, Universidad de Tarapac\'{a}, Casilla 7D, Arica, Chile
\and
Instituto de Astronom\'{i}a Te\'{o}rica y Experimental, (IATE, CONICET-UNC), C\'{o}rdoba, Argentina
\and
Universidad Nacional de C\'{o}rdoba, Observatorio Astron\'{o}mico de C\'{o}rdoba
\and
Department of Astronomy \& Astrophysics, The Pennsylvania State University, University Park, PA 16802, USA
\and
The Institute for Gravitation and the Cosmos, The Pennsylvania State University, University Park, PA 16802, USA
\and
Department of Physics, 104 Davey Laboratory, The Pennsylvania State University, University Park, PA 16802, USA
\and
Department of Physics, University of Napoli ``Federico II'', Via Cinthia 9, 80126 Napoli, Italy
\and
INAF -- Osservatorio Astronomico di Capodimonte, Via Moiariello 16, 80131 Napoli, Italy
\and
The Graduate Center of the City University of New York, 365 Fifth Avenue, New York, NY 10016, USA
\and
Department of Astrophysics, American Museum of Natural History, Central Park West and 79th Street, NY 10024-5192, USA
\and
Department of Physics and Astronomy, Lehman College of the CUNY, Bronx, NY 10468, USA
\and
Ru{\dj}er Bo{\v s}kovi{\'c} Institute, Bijeni{\v c}ka Cesta 54, 10000 Zagreb, Croatia
\and
Centre for Astrophysics and Supercomputing, Swinburne University of Technology, Hawthorn, VIC 3122, Australia
\and
Shanghai Astronomical Observatory, Chinese Academy of Sciences, 80 Nandan Road, Shanghai 200030, People's Republic of China
\and
Centro Interdisciplinario de Data Science, Facultad de Ingenier\'{i}a y Ciencias, Universidad Diego Portales, Av. Ej\'{e}rcito Libertador 441, Santiago, Chile
\and
Department of Astronomy, Faculty of Mathematics, University of Belgrade, Studentski trg 16, 11000 Belgrade, Serbia
\and
Hamburger Sternwarte, Universit\"at Hamburg, Gojenbergsweg 112, D-21029 Hamburg, Germany
\and
Departamento de Astronom\'{i}a, Universidad de Chile, Camino el Observatorio 1515, Santiago, Chile
\and
Space Telescope Science Institute, 3700 San Martin Dr., Baltimore, MD 21218, USA
\and
Universidade Federal de Santa Maria (UFSM), Centro de Ci{\^e}ncias Naturais e Exatas (CCNE), Santa Maria, 97105-900, RS, Brazil
\and
Laborat\'{o}rio Interinstitucional de e-Astronomia - LIneA, Rua Gal. Jos\'{e} Cristino 77, Rio de Janeiro, RJ - 20921-400, Brazil
\and
International Gemini Observatory/NSF NOIRLab, Casilla 603, La Serena, Chile
\and
NASA Goddard Space Flight Center, Greenbelt, MD 20771, USA
\and
Center for Space Science and Technology, University of Maryland Baltimore County, MD 21250, USA
\and
Department of Astronomy, University of Geneva, ch. d'Ecogia 16, 1290, Versoix, Switzerland
\and
Department of Physics, Drexel University, 32 S. 32nd Street, Philadelphia, PA 19104, USA
\and
Departamento de F\'{i}sica, Universidad T\'{e}cnica Federico Santa Mar\'{i}a, Vicu{\~n}a Mackenna 3939, San Joaqu\'{i}n, Santiago, Chile
\and
School of Physics \& Astronomy, University of Southampton, Highfield Campus, Southampton SO17 1BJ, UK
\and
Centre for Extragalactic Astronomy, Department of Physics, Durham University, South Road, Durham DH1 3LE, United Kingdom
\and
National Radio Astronomy Observatory, 520 Edgemont Road, Charlottesville, VA 22904, USA
\and
Department of Astronomy, University of Virginia, 530 McCormick Rd, Charlottesville, VA 22904, USA
\and
Department of Astronomy, University of Michigan, 1085 S University, Ann Arbor, MI 48109, USA
}
\authorrunning{Satheesh-Sheeba et al.}
\date{Accepted XXX. Received YYY; in original form ZZZ}
\abstract{
Photometric redshift estimation for active galactic nuclei (AGNs) remains a fundamental challenge for current and upcoming large-scale photometric surveys. Traditional spectral energy distribution (SED) fitting suffers from color-redshift degeneracies, particularly for AGNs whose power-law continua hide the strong spectral features required to anchor redshift estimates. While AGN variability has been shown to provide additional constraining power, existing frameworks require multi-band light curves that are not always available. This work presents \texttt{VAR-PZnn}, a fully connected mixture density network that integrates 26 variability features extracted from ZTF \textit{g}-band light curves with optical photometry from Pan-STARRS1, mid-infrared (MIR) photometry from CatWISE, and, for a subsample, near-infrared (NIR) photometry from UKIDSS. The model is trained and tested on 72\,728 spectroscopically confirmed AGNs/QSOs spanning a redshift range of $0.01<z<4.5$ and \textit{g}-band magnitudes from 17 to 21.5. For the main sample, we achieve a normalized median absolute deviation of $\sigma_{\mathrm{NMAD}}=0.058$ and a catastrophic outlier fraction of $\eta=8.2\%$, which is reduced to $5.4\%$ when the 10\% of sources with the highest predicted uncertainty are excluded. A systematic ablation study demonstrates that MIR photometry provides the dominant constraint for photometric redshift (photo-$z$) accuracy, while variability features serve as a secondary refiner. Using UKIDSS NIR data as a proxy for future synergies between LSST and space-based NIR missions like \textit{Euclid} and \textit{Roman}, we obtain additional results with $\eta=13.3\%$ without MIR data and $\eta=4.6\%$ when MIR photometry is available. We benchmark our model against Low-Resolution Templates (\texttt{LRT}) SED template-fitting ($\eta=28.7\%$) and the \texttt{VAR-PZ} variability-based framework; applying single-band \texttt{VAR-PZ} priors worsens the \texttt{LRT} performance to $\eta=39.4\%$ due to degenerate solutions inherent to single-band light curves. This interpretation is confirmed through simulations of ZTF-like light curves, where the degradation is only marginal ($\eta=27.6\%$ to $28.1\%$). These results demonstrate that combining time-domain variability with multi-wavelength photometry in a neural-network framework provides a robust and scalable approach for AGN photo-$z$ estimation in preparation for the Vera C. Rubin Observatory's Legacy Survey of Space and Time.
}


\maketitle

\section{Introduction}

It is widely accepted that active galactic nuclei (AGNs) are powered by the accretion of matter onto their central supermassive black holes (SMBH; \citealt{Lyndenbell1969}). This process releases energy across the entire electromagnetic spectrum from physically distinct components, including the accretion disk, the X-ray corona, the torus, or relativistic jets \citep{Padovani2017,Paolillo2025}. Due to their high luminosities, AGNs can be observed across cosmic time, providing a probe to study black-hole growth and its connection to galaxy evolution \citep{Kormendy2013,Madau_2015}. Recent studies have clarified that the coevolution of SMBHs and their host galaxies is a complex, non-linear process shaped significantly by galactic mergers 
\citep[e.g.,][]{Graham_2023,sahu_graham_2023}. To fully leverage these observations and understand AGN feedback, its regulatory role in star formation \citep{Silk_Rees,Fabian_2012}, and the coevolution of SMBHs and host galaxies \citep [e.g.,][]{Kormendy2013,Brandt_alexander2015,brandt_yang2022}, accurate redshifts are necessary. Large-scale photometric surveys, such as the Vera C. Rubin Observatory's Legacy Survey of Space and Time (LSST; \citealt{ivezic2019}), are expected to observe over 10 million quasars and over 100 million AGNs \citep{lsstsciencebook,Li_2025}. While current and upcoming
spectroscopic surveys such as 4MOST \citep{DeJong2019}, DESI \citep{DESI2016}, SDSS \citep{York2000,Kollmeier_2026}, and MOONS \citep{Cirasuolo_2012} will observe a fraction of these sources, the vast majority will lack spectroscopy \citep{Dahlen2013}. This limitation will only compound with space-based missions such as \textit{Euclid} \citep{Euclid_study_report} and \textit{Nancy Grace Roman Space Telescope} \citep{roman2024}, which will detect new AGNs through deep imaging that falls well below the sensitivity limits of available spectroscopic follow-up \citep{Euclid}. Consequently, photometric redshift (photo-$z$) estimation using multi-band photometry is required for the identification of the bulk of the AGN population.

 Spectral energy distribution (SED) fitting is widely used as a standard method for estimating photo-$z$s. While tools such as \texttt{LePHARE} were originally created for normal galaxies \citep[e.g.,][]{Arnouts1999,Ilbert2006}, they have since been successfully tuned and optimized for AGN populations \citep[e.g.,][]{salvato2011dissecting,salvato2019many,salvato2022erosita,Shirley_submitted,Ananna2017}. However, extracting accurate photo-$z$s from SED modeling presents challenges.
 The power-law continuum of AGNs lacks the strong spectral features required to anchor redshift estimates. Additionally, non-simultaneous observations and intrinsic variability distort observed broad-band fluxes. Resolving these degeneracies in SED fitting requires dense photometric coverage \citep{Brescia2019,Saxena_2024} or strict priors, such as redshift-dependent absolute magnitude constraints \citep{Fotopoulou2012,Hsu2014,Ananna2017,Peca2021}. 

In the current era of large-scale photometric surveys, machine learning (ML) offers a computationally efficient alternative for large-scale data analysis. Recent studies have applied ML algorithms for AGN selection \citep[e.g.,][]{Savic2023,sanchez_ZTF_2023,De_Cicco2025,arevalo2025,Laura2026} and photo-$z$ estimation \citep[e.g.,][] {Saxena_2024,Roster2024} using multi-band photometry and image features. Because ML models are capable of capturing non-linear features, they generally achieve higher accuracy for sources well-represented in the training data. However, data-driven approaches produce a higher fraction of catastrophic outliers when applied to sources outside the training sample parameter space \citep{Duncan2018,salvato2022erosita}. Overcoming these degeneracies requires incorporating additional physical constraints, such as time-domain variability. The extraction of such variability constraints from wide-field surveys, such as the Zwicky Transient Facility (ZTF; \citealt{ZTF}) and the current LSST, has been facilitated by community brokers like the Automatic Learning for the Rapid Classification of Events (ALeRCE; \citealt{Alerce}) \footnote{\url{https://alerce.science}}. A core component of the ALeRCE architecture is its feature extraction pipeline, which computes dozens of metrics from irregularly sampled optical light curves to characterize the photometric behavior of transient and variable sources \citep[hereafter \citetalias{alerts_classifier}]{alerts_classifier}. These extracted features capture essential temporal dynamics through basic statistical properties, periodicity indicators, and stochastic modeling. 

For AGNs, which typically exhibit stochastic variability, light-curve features serve as physical tracers. Parameters derived from damped random walk (DRW) models \citep{Kelly2009,MacLeod_2010,Suberlak_2021}, continuous-time auto-regressive moving average (CARMA) processes \citep{CARMA}, as well as the Mexican Hat Power Spectrum \citep{mexican_hat,Lorena2026} effectively capture the characteristic timescale and amplitude of accretion-driven variability \citep[hereafter \citetalias{sanchez_ZTF_2023}]{sanchez_ZTF_2023}. Recent studies demonstrate that these temporal features are also correlated with the underlying physical properties of the SMBH and its accretion disk \citep{SanchezSaez2021b,Burke2021,arevalo2023universalpowerspectrumquasars,Paolillo2025,Petrecca2026}. Additionally, the observed AGN variability is redshift-dependent \citep{Saxena_2024,VARPZ} as cosmological time dilation stretches the observed timescales by a factor of (1+z) and therefore modeling variability can provide additional constraints for photo-$z$ estimation, as demonstrated by the \texttt{VAR-PZ} framework \citep{VARPZ}. The original \texttt{VAR-PZ} method calculates a variability-based posterior probability using theoretical scaling relations between DRW parameters and redshift, which is then multiplied as an independent prior onto traditional SED-fitting templates. That study demonstrated that AGN variability is a useful tool, capable of improving the photo-$z$ outlier fractions ($\eta$) by 10\% using SDSS light curves, and improving them by 25\% when applied to simulated 10-year LSST light curves. However, the extent to which single-band variability can resolve color-redshift degeneracies when extensive multi-wavelength photometry is already available remains an open question.

In this work, we investigate whether time-domain variability provides a measurable improvement to AGN photometric redshift estimation once multi-wavelength photometry is already available. To this end, we present \texttt{VAR-PZnn}, a fully-connected mixture density neural network that combines variability features extracted from ZTF \textit{g}-band light curves with optical photometry from Pan-STARRS1, mid-infrared (MIR) photometry from CatWISE, and, for a subsample, near-infrared (NIR) photometry from UKIDSS. The variability features were compiled by \citet{arevalo2025} (hereafter \citetalias{arevalo2025}) and are adapted from the metrics defined by the ALeRCE broker for light-curve classification \citepalias{sanchez_ZTF_2023}. We use ZTF as a test bed for Rubin-LSST-like time-domain data and quantify the relative contribution of variability, optical colors, MIR photometry, and NIR photometry to the final photo-$z$ performance. In addition, by analyzing the UKIDSS subsample, we assess the extent to which optical and NIR information can compensate for the absence of MIR data, thereby providing a useful proxy for future Rubin-\textit{Euclid}-\textit{Roman} AGN photo-$z$ applications.

The article is structured as follows. Section \ref{data} describes the datasets used in this study, detailing the adopted ZTF variability metrics alongside the corresponding optical photometry from Pan-STARRS, NIR photometry from UKIDSS Large Area Survey (LAS), mid-infrared (MIR) from CatWISE, and spectroscopic data from SDSS and DESI. Section \ref{Features} describes our sample selection, outlining the construction of our training and testing sets and the features used for our analysis. In Section \ref{Method}, we present our methodology, detailing the neural network architecture used to integrate these multi-wavelength and time-domain features, as well as the evaluation metrics used to quantify the performance of our algorithm. Section~\ref{results} presents our results, specifically, we quantify the photo-$z$ precision and the outlier fraction for two distinct cases: a primary sample incorporating both optical and MIR photometry, and a sub-sample using optical and NIR data as a proxy to forecast the future synergies of LSST with \textit{Euclid} and \textit{Roman} missions. We also benchmark our results against traditional SED modeling methods and the \texttt{VAR-PZ} framework. Finally, we summarize our main conclusions in Section \ref{conclusions}. Throughout this paper, the magnitudes are reported in the AB system \citep{1983Oke}.

\section{Data}\label{data}

\subsection{ZTF variability data}

The parent catalog compiled by \citetalias{arevalo2025} contains approximately 451\,497 variable sources. These sources were initially selected and classified using a random forest light-curve classifier \citepalias{arevalo2025}, which relies on time-domain and color features to distinguish between stochastic (e.g., AGN, QSOs), periodic, and transient phenomena. For our analysis, we restricted this sample to only those objects classified as either \texttt{AGN} or \texttt{QSO}, which reduced the sample to 83\,453 sources. Utilizing a sample that is fundamentally pre-selected based on variability is highly relevant for forecasting LSST capabilities, as LSST's primary method for identifying new AGN populations, prior to any spectroscopic follow up, will rely on exactly this type of automated time domain classification. The variability parameters for these were extracted from custom ZTF forced photometry using difference-image aperture photometry (DI-Ap). Unlike standard point spread function (PSF) fitting techniques, the DI-Ap approach is optimized to decouple the intrinsic AGN signal from the host-galaxy light, whose apparent contribution can otherwise vary with seeing fluctuations. These measurements were computed only for sources within the sky region relevant to the overlap between ZTF and the 4MOST survey, restricted to $-29^\circ < \delta_{J2000} < +15^\circ$, where $-29^\circ$ corresponds to the southern declination limit of ZTF, and to Galactic latitudes $|b| > 20^\circ$ to avoid the Galactic plane. The light curves used to derive these quantities extend up to September 2022, corresponding approximately to ZTF DR13. These variability quantities were derived from the ZTF \textit{g}-band light curves, and therefore represent a single-band characterization of the source variability.  Consequently, this imposes a natural limit on the spectroscopic redshift coverage of our sample, as the requirement for \textit{g}-band detections inherently excludes sources at high redshifts where the flux is strongly attenuated by intergalactic medium (IGM) absorption \citep{Madau1995,Inoue2014}.

\subsection{Photometric data}

For the ancillary photometric information, we adopted the catalog compiled by \citetalias{arevalo2025}. This catalog includes, when available, optical photometry in the $g$, $r$, $i$ and $z$ bands from Pan-STARRS1 (PS1, \citealt{PS1}), together with MIR photometry in the W1 and W2 bands from CatWISE \citep{Eisenhardt_2020,Marocco_2021}. The longer-wavelength W3 and W4 bands are not included in this analysis because their significantly shallower depth leads to lower detection fractions, which would severely reduce the completeness of the cross-matched sample. As described in \citetalias{sanchez_ZTF_2023}, counterparts to the ZTF sources were identified through positional matching using radii of $1.5^{\prime\prime}$ for PS1 (using magnitudes measured with PSF photometry) and $2^{\prime\prime}$ for CatWISE (using magnitudes extracted through profile-fitting photometry). All photometric measurements were corrected for Galactic extinction prior to our analysis \citep{Schlegel1998}. We note that the photometric measurements used in this work are not simultaneous, having been obtained at different epochs. For variable AGN, this non-simultaneity can introduce effective color offsets that do not correspond to any physical SED state. While this effect is partially mitigated by the use of time-averaged (stacked) photometry in both surveys, it remains a potential source of systematic error, particularly for highly variable sources.

\subsection{Spectroscopic catalog}

The spectroscopic redshifts used in this work were compiled by cross-matching the \citetalias{arevalo2025} sample with public AGN catalogs from the sixteenth and nineteenth data releases of the Sloan survey (SDSS DR16/DR19, \citealp{Lyke2020,SDSS_DR19}) and data release 1 (DR1) of the DESI survey \citep{DESI_DR1} using a $1''$ matching radius. These spectroscopic redshifts are used as the reference labels for our supervised ML analysis and for the evaluation of the model performance.

After cross matching the \citetalias{arevalo2025} \texttt{AGN/QSO} sample with the available spectroscopic information, we obtained 74\,331 sources with spectroscopic redshifts. After excluding sources with invalid or missing photometric data, our final sample comprises 72\,728 valid sources, which define the labeled spectroscopic sample used in this work.

\subsection{Near-infrared photometry from UKIDSS}
 
In addition to the optical and MIR photometry, we also used the NIR photometry from the UKIRT Infrared Deep Sky Survey \citep[UKIDSS;][]{UKIDSS}. UKIDSS was carried out with the Wide Field Camera at the 3.8 m United Kingdom Infrared Telescope (UKIRT). We used the data release 11 plus catalog from the Large Area Survey (LAS), accessed through the NOIRLab Astro Data Lab TAP service\footnote{\url{https://datalab.noirlab.edu/data/ukidss}}.

By cross-matching our spectroscopic AGN sample with the UKIDSS LAS catalog, we obtained 47\,906 sources with $Y$, $J$, $H$, and $K$ band aperture photometry. We used these measurements to derive additional NIR colors for the analysis. These NIR colors were included as an approximation to obtain a first estimate of the improvement that can be expected from the LSST-\textit{Euclid} synergy \citep{Euclid}. It is important to note that while UKIDSS includes the $K$ band, the \textit{Euclid} NISP instrument only covers the $Y_E$, $J_E$, and $H_E$ bands. Furthermore, the \textit{Euclid} filters are broader than their UKIDSS counterparts and have no gaps in wavelength coverage between them. Therefore, this UKIDSS-based proxy serves as an exploratory first estimate. This analysis is particularly useful for sources without WISE data, for which the UKIDSS-based NIR colors provide an alternative constraint.

\begin{figure}[h]
    \centering
    \includegraphics[width=0.95\linewidth]{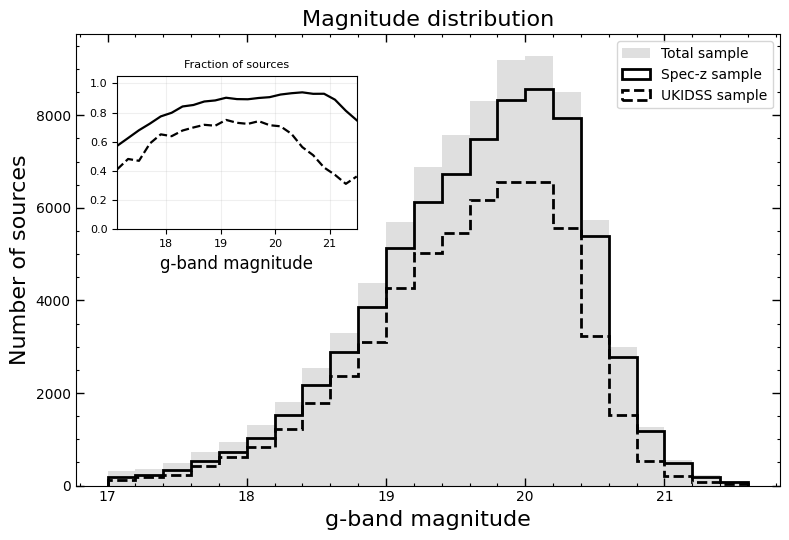}
     \caption{Distribution of the samples used in our analysis as a function of brightness. The grey histogram represents the 83\,453 sources from the parent \citetalias{arevalo2025} catalog explicitly classified as AGN or QSO. The inset shows the fraction of these sources with spectroscopic redshifts and with UKIDSS detections (used in our study).}
    \label{fig:magnitude_distribution}
\end{figure}

Figure \ref{fig:magnitude_distribution} shows the distribution of the sample as a function of their \textit{g}-band magnitudes, used in our study. The grey-colored histogram represents the \citetalias{arevalo2025} sample which is the total number of sources classified as \texttt{AGN/QSO}, the black histogram shows the sample with spectroscopically confirmed redshifts from SDSS/DESI and the dashed line represents the sources with NIR measurements from UKIDSS. The inset depicts the fraction of the spectroscopic sample and the UKIDSS sample compared to the total sample.

\section{Sampling and features}\label{Features}

We used the labeled sample of 72\,728 sources for our analysis and randomly split it into training, validation, and test sets in a 60:20:20 ratio, respectively. We also used a subsample of 47\,906 sources with UKIDSS photometry as an additional test, for which the same training, validation, and test partitions were adopted. 
\begin{figure}[h]
    \centering
    \includegraphics[width=0.95\linewidth]{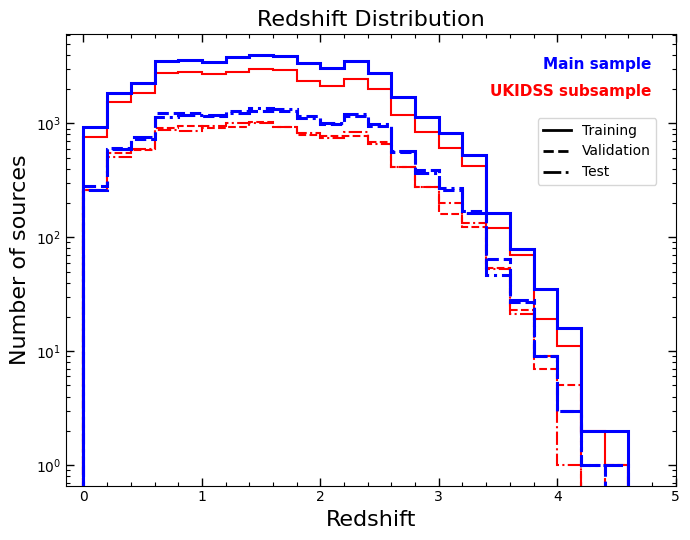}
    \caption{Spectroscopic redshift distributions of the training, validation, and test samples adopted in our analysis, represented by solid, dashed, and dotted lines, respectively. Blue denotes the main sample, comprising 72\,728 sources, while red corresponds to the UKIDSS subsample, containing 47\,906 sources.}
    \label{fig:redshift_distribution}
\end{figure}
Figure \ref{fig:redshift_distribution} illustrates the redshift distributions of the sources used in our analysis with the distributions for both the main sample (72\,728 sources, shown in blue) and the UKIDSS subsample (indicated in red). The targets span a wide redshift range, with the bulk of the population peaking broadly between $z\sim1.0$ and $z\sim2.5$, which is characteristic of optically selected AGN catalogs \citep{Richards2002}. The figure illustrates that the redshift distributions of the validation and test sets closely track the training distribution across the entire redshift range. This consistency ensures that our model is trained and evaluated on a representative sample of the parameter space, avoiding redshift-dependent sampling biases.

The input features include optical and mid-infrared colors, together with variability features derived from the optical light-curve analysis. The complete list of adopted features and their references are given in Table \ref{appendix:features_table}.

We adopted a set of 26 variability features extracted from ZTF light curves by \citetalias{arevalo2025}, following the prescription given in \citetalias{alerts_classifier} and \citetalias{sanchez_ZTF_2023}. These features characterize different variability properties of the sources, including amplitude, characteristic timescale, light-curve distribution, and stochastic behavior. In addition to the variability features, we incorporated color and morphology information derived from PS1, CatWISE, and UKIDSS photometry. For the main sample, we used the optical colors from PS1 $(griz)$ bands, optical-MIR colors combining PS1 and CatWISE $(W1, W2)$ bands, and the MIR color $W1-W2$. 
For the subsample with available UKIDSS photometry, we also incorporated NIR color information derived from the $Y$, $J$, $H$ and $K$ bands. The photometric dataset corresponds to 16 optical-NIR colors, 6 NIR-MIR colors and 6 NIR colors. Furthermore, we included the PS1 morphology flag from \citealt{Tachibana2018} as an additional input feature.

\section{Methodology} \label{Method}

We implemented a supervised regression model based on a Mixture Density Network (MDN; \citealt{mixed_density_network}) for training and photo-$z$ estimation. The input feature vector is composed of the variability features described in Section~\ref{Features}, together with the color, magnitude, and morphology features.

\subsection{Input sample and feature preparation}
\label{subsec:input_features}

The training table contains one row per source, with the spec-$z$ being the true label. We constructed the final feature matrix by combining the 26 variability features with the magnitudes, colors, and the PS1 morphology flag (see Section \ref{Features}).

Because the variability features span a wide dynamic range and can show non-Gaussian distributions, we first applied a quantile-based transformation to the variability features. Specifically, if $x_j$ denotes the $j^{\mathrm{th}}$ variability feature, for each feature, we mapped its empirical cumulative distribution to a Gaussian target distribution,
\begin{equation}
x'_j = \Phi^{-1}\!\left[F_j(x_j)\right],
\end{equation}
where $F_j$ is the empirical cumulative distribution function of the feature and $\Phi^{-1}$ is the inverse cumulative distribution function of a standard normal distribution. We emphasize that this is a mathematical transformation applied directly to the feature values to map them to a normal distribution; no sources are discarded or cut from the sample during this process. This step normalizes highly skewed variability descriptors and reduces the impact of long tails before training. This transformation is an additional pre-processing step applied prior to the neural-network training.

After this step, all input features were standardized using the mean and standard deviation computed from the training subset,
\begin{equation}
\tilde{x^i_j} = \frac{x^i_{j} - \mu_j}{\sigma_j},
\end{equation}
where $x^i_{j}$ is the value of feature $j$ for source $i$, and $\mu_j$ and $\sigma_j$ are the mean and standard deviation of that feature in the training sample. The same transformation was then applied to the validation and test subsets.

We note that individual measurement uncertainties (e.g., photometric errors on magnitudes, or uncertainties on the variability descriptors) are not provided as additional input features to the network. This is a deliberate design choice; the rationale and implications are discussed in Section~\ref{caveats}.

\subsection{Training-validation-test split}
\label{subsec:data_split}

As mentioned before, we divided the sample randomly into training, validation, and test subsets using a 60:20:20 split, with no overlap between the subsets. Specifically, 60\% of the objects were used as the training sample, while the remaining 40\% were split equally into validation and test samples. The training sample was used to optimize the network weights, the validation sample was used to monitor convergence and select the best model, and the test sample was kept aside for the final performance assessment. 

\subsection{Network architecture}
\label{subsec:network_architecture}

A standard artificial neural network (ANN) trained for regression outputs a single point estimate of the redshift for each source and provides no built-in measure of the prediction uncertainty. This is a significant limitation for photo-$z$ estimation, where the mapping between broadband colors and redshift can be inherently multi-valued: different redshifts can produce nearly identical photometric colors, leading to degenerate solutions that a single point estimate cannot capture. To address this, we adopt a Mixture Density Network \citep[MDN;][]{mixed_density_network}, which replaces the single-valued output of a standard ANN with the parameters of a Gaussian mixture model. Instead of predicting a single redshift, the MDN outputs, for each source, the full set of mixture weights, means, and standard deviations that define a flexible probability distribution over redshift. This formulation naturally captures multi-modal posteriors and provides well-defined predictive uncertainties as a direct output of the network.

Specifically, we adopted an MDN with three hidden layers of 128 neurons each, using rectified linear unit (ReLU) activations and dropout rate of $0.2$ for
regularization. The output layer parameterizes a mixture of $K=3$ Gaussian 
components (see Appendix ~\ref{appendix:Gaussian} for a performance comparison of different $K$ values), predicting for each source the mixing coefficients 
$\{\pi_k\}$, mean $\{\mu_k\}$, and standard deviations $\{\sigma_k\}$ 
($k=1,\ldots,K$). The predicted photo-$z$ is taken as the expected value 
of the mixture:
\begin{equation} \label{Eq3}
  \hat{z} = \sum_{k=1}^{K} \pi_k \, \mu_k.
\end{equation}
The network was trained by minimizing the negative log-likelihood (NLL) of 
the Gaussian mixture model:
\begin{equation}
  \mathcal{L}_{\rm NLL} = -\frac{1}{N}\sum_{i=1}^{N} 
  \log\left[\sum_{k=1}^{K} \pi_k^{(i)} \, 
  \mathcal{N}(z_{\mathrm{spec},i} \mid \mu_k^{(i)}, \sigma_k^{(i)})\right].
\end{equation}
where $N$ is the number of objects in a given training batch. Optimization was performed with the Adam optimizer \citep{kingma2014adam}, using a learning rate of $10^{-4}$ and an $L_2$ weight decay of $10^{-4}$. To ensure stability during the MDN training, we applied gradient clipping with a maximum norm of 1.0. We adopted a batch size of 256 and trained for a maximum of 250 epochs. During training, the validation loss was monitored after each epoch, and early stopping was applied with a patience of 20 epochs to prevent overfitting.

\subsection{Uncertainty estimation}
\label{subsec:uncertainty}

While the MDN naturally provides a full posterior distribution $p(z \mid \mathbf{x}) = \sum_k \pi_k \mathcal{N}(z \mid \mu_k, \sigma_k)$ for each source, practical catalog applications typically require a scalar point estimate and associated uncertainty. To estimate the summarized photo-$z$ uncertainties of our predictions, we follow the framework of \citet{Kendall2017} and decompose the total predictive variance into aleatoric and epistemic components via Monte Carlo (MC) Dropout \citep{gal16}. We perform $T=50$ stochastic forward passes with dropout enabled at inference time.

For each forward pass $t$, the \textit{aleatoric uncertainty}, which captures the intrinsic noise in the 
data and the multi-modal nature of the color-redshift mapping, is obtained 
directly from the MDN output:
\begin{equation}
    \sigma^{2\,(t)}_{\rm aleatoric} = \sum_{k=1}^{K} \pi_k^{(t)} 
  \left(\sigma_k^{2\,(t)} + \mu_k^{2\,(t)}\right) - \hat{z}^{2\,(t)}.
\end{equation}

The \textit{epistemic uncertainty}, which reflects the model's lack of 
knowledge due to finite training data, is estimated from the variance 
of the expected means across passes:
\begin{equation}
  \sigma^2_{\rm epistemic} = \mathrm{Var}_{t=1}^{T}
  \left[\hat{z}^{(t)}\right].
\end{equation}

The total predictive uncertainty is then obtained by averaging the per-pass aleatoric variances and adding the epistemic variance \citep[Eq.~9 of][]{Kendall2017}:
\begin{equation}
\sigma_z = \sqrt{\frac{1}{T}\sum_{t=1}^{T}\sigma^{2\,(t)}_{\rm aleatoric} + \sigma^2_{\rm epistemic}}.
\end{equation}

\subsection{Performance metrics}
\label{subsec:metrics}

The performance of the network was evaluated by comparing the predicted photo-$z$s with the spec-$z$s. We define the normalized redshift residual as
\begin{equation}
\Delta z_{\rm norm} =
\frac{z_{\rm phot} - z_{\rm spec}}{1+z_{\rm spec}}.
\end{equation}

Following \citet{salvato2019many}, we computed the outlier fraction ($\eta$) as the fraction of sources satisfying
$\left|\Delta z_{\rm norm}\right| > 0.15$
and we quantified the scatter using the normalized median absolute deviation (NMAD),
\begin{equation}
\sigma_{\rm NMAD} =
1.48 \times {\rm median}
\left(
\left|
\frac{z_{\rm phot} - z_{\rm spec}}{1+z_{\rm spec}}
\right|
\right).
\end{equation}

Finally, we define the root mean square error (RMSE), which serves as the baseline metric for evaluating the permutation feature importance (see Section. \ref{subsec:diagnostics}):
\begin{equation}
{\rm RMSE} =
\sqrt{
\frac{1}{N}
\sum_{i=1}^{N}
\left(\hat{z}_i - z_{{\rm spec},i}\right)^2
}.
\end{equation}

\noindent where $\hat{z}_i$ is the predicted photo-$z$ from Eq \ref{Eq3}. We note that, unlike $\sigma_{\rm NMAD}$ and $\eta$, this metric is not normalized by $(1+z_{\rm spec})$. Since $\Delta{\rm RMSE} = {\rm RMSE}^{\rm perm}j - {\rm RMSE}{\rm ref}$ is used only for the relative ranking of features, and the same metric is applied consistently across all features, the absence of the $(1+z_{\rm spec})$ normalization does not affect the resulting ranking.

\subsection{Diagnostic plots and feature-importance analysis}
\label{subsec:diagnostics}

To quantify the training behavior and the predictive performance of the model, we generated a set of diagnostic products for each experiment: (i) the training and validation loss curves as a function of epochs; (ii) the comparison between $z_{\rm phot}$ and $z_{\rm spec}$ for the test sample; (iii) the redshift distributions of the training, validation, and test subsets; and (iv) the redshift-dependent evolution of the NMAD and outlier fraction in bins of width $\Delta z = 0.5$.

Beyond assessing overall accuracy, it is crucial to understand which photometric inputs drive the network's predictions. To quantify the relative contribution of each input variable, we evaluated the permutation feature importance \citep{Breiman2001,fisher2019}. This technique measures the drop in model performance when the information from a single feature is effectively removed. We define the importance $I_j$ of a given feature $j$ as:

$$I_j = {\rm RMSE}^{\rm perm}_j - {\rm RMSE}_{\rm ref},$$

\noindent where ${\rm RMSE}_{\rm ref}$ is the baseline root-mean-square error evaluated on the uncorrupted test set, and ${\rm RMSE}^{\rm perm}_j$ is the RMSE calculated after randomly shuffling the values of feature $j$ across all objects in the test sample. By randomly permuting the values of a single feature among the different instances, we break its true physical association with the target redshift while preserving its overall statistical distribution. Consequently, features producing the largest increase in RMSE (i.e., the highest $I_j$) were interpreted as the most influential for the redshift prediction. Figure \ref{fig:flowchart} summarizes the workflow of our neural-network framework, including feature input, pre-processing, training, validation, and photo-$z$ predictions.

\section{Results}\label{results}
\subsection{Application to the main sample}\label{main sample}

We trained our model on 43\,618 sources ($\sim60\%$ of the main sample), validated it on 14\,555 sources ($\sim20\%$), and used the remaining 14\,555 sources ($\sim20\%$) as an independent test set.
\begin{figure}
    \centering
    \includegraphics[width=0.9\linewidth]{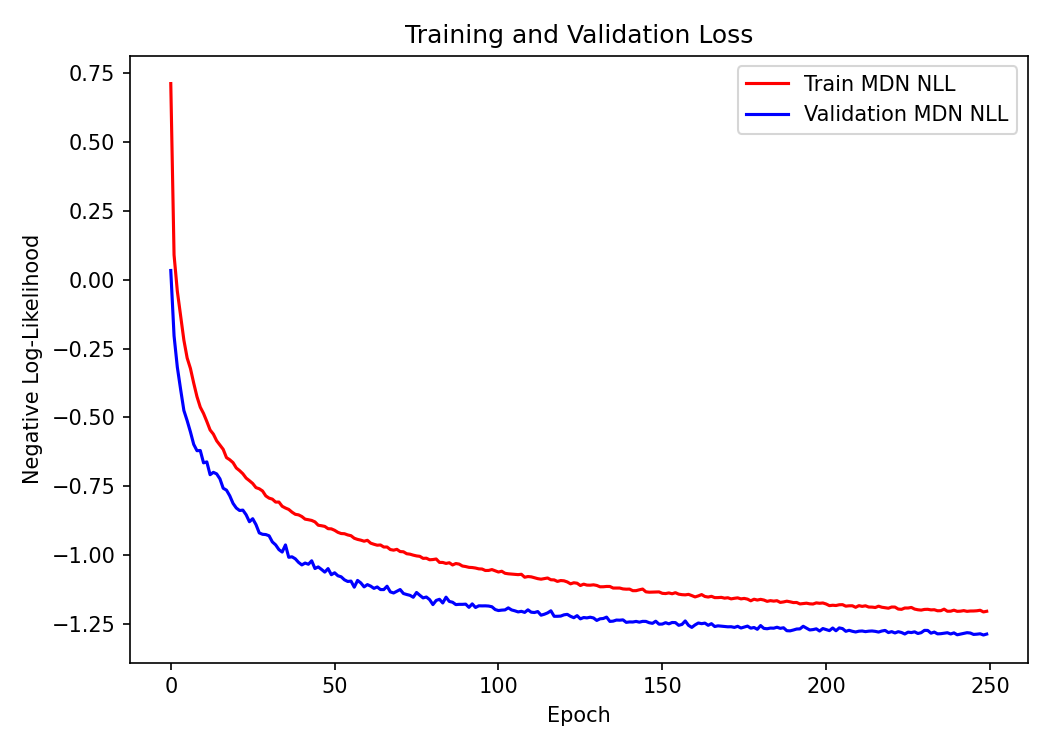}
    \caption{Evolution of the loss function with training epoch for the training (red solid line) and validation (blue solid line) sets.}
    \label{fig:lossfunction_mainsample}
\end{figure}
Figure \ref{fig:lossfunction_mainsample} shows the evolution of the NLL loss function during the training and validation steps of our model. Both curves decrease rapidly during the first few epochs and then gradually flatten, indicating that the network quickly learns the main structure of the data before slowly approaching convergence. The training and validation losses remain closely aligned throughout the process, with the validation loss slightly below the training loss, suggesting stable learning and no clear evidence of overfitting.
\begin{figure}
    \centering
    \includegraphics[width=1\linewidth]{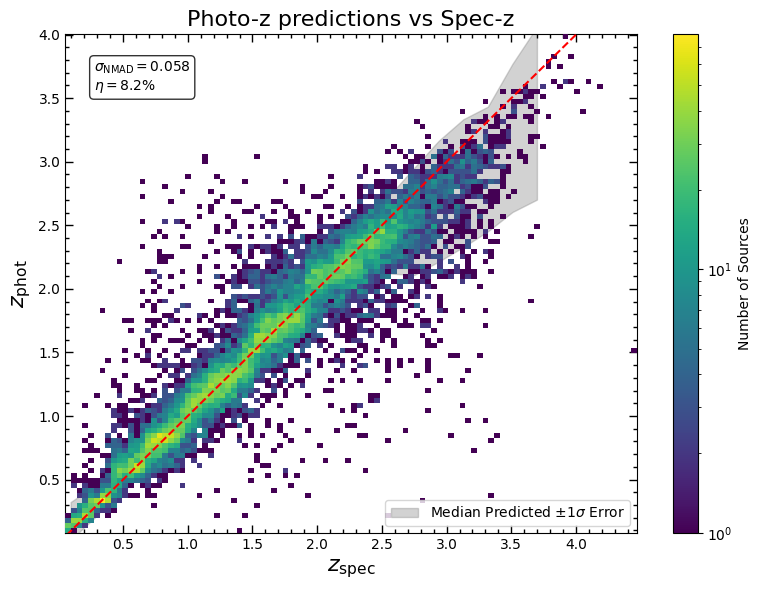}
    \caption{A comparison of the photo-$z$ predictions with 
their spectroscopic redshifts for our test sample. The red dashed line represents the 
one-to-one relation. The shaded grey region indicates the median 
predicted $\pm 1\sigma$ uncertainty envelope. The corresponding 
$\sigma_{\mathrm{NMAD}}$ and $\eta$ are shown in 
the inset.}
    \label{fig:photoz_predictions_mainsample}
\end{figure}
\begin{figure}
    \centering
    \includegraphics[width=1\linewidth]{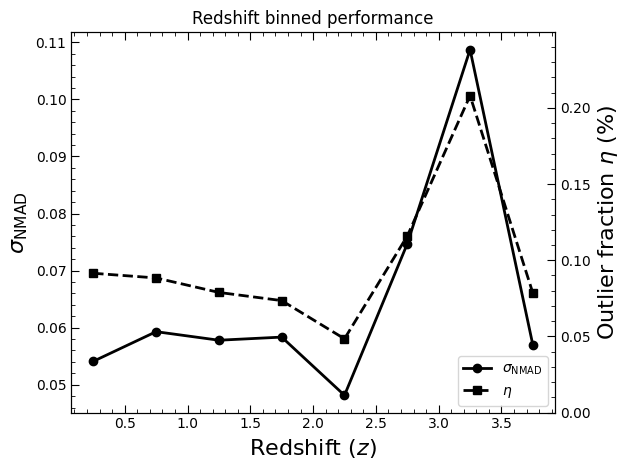}
    \caption{Photo-$z$ performance as a function of redshift bin. The solid line with circles shows the $\sigma_{\mathrm{NMAD}}$, while the dashed line with squares indicates the $\eta$ in percent.}
    \label{fig:redshift_binned_photo_z_mainsample}
\end{figure}
Figure \ref{fig:photoz_predictions_mainsample} compares the spectroscopic redshifts with the predicted photo-$z$s. The predicted photo-$z$s are in agreement with their true redshifts, with the majority of the sources lying close to the one-to-one relation. The source density along that relation demonstrates that the model is able to recover the redshifts reliably over most of the sampled parameter space. The overall photo-$z$ predictions yield a $\sigma_{\mathrm{NMAD}}$ of 0.058 and an $\eta$ of 8.2\%. While normal inactive galaxies allow for highly precise photo-$z$s ($\sigma_{\mathrm{NMAD}}\sim0.01-0.02$; e.g., \citealt{Ilbert2009, Laigle2016L}), achieving stable predictions for active galaxies remains a recognized challenge. Given this disparity, our results demonstrate that our algorithm yields robust predictions for AGN photo-$z$s. 
The shaded grey region in Figure \ref{fig:photoz_predictions_mainsample} indicates the median predicted $\pm 1\sigma$ uncertainty envelope. This contour closely traces the underlying scatter of the predictions, naturally widening at higher redshifts where the model's confidence decreases. This alignment demonstrates that the predicted uncertainties ($\sigma_z$) serve as a reliable diagnostic for the photo-$z$ quality. By using this confidence metric to filter the sample, we find that excluding the 10\% of sources with the highest predicted uncertainty significantly improves the reliability of the catalog, reducing the outlier fraction from 8.2\% to 5.4\%. A quantitative assessment of the uncertainty calibration is presented in Appendix~\ref{appendix:calibration}, where we show through a coverage analysis that these predicted uncertainties are slightly conservative, systematically over-covering the true scatter (mean calibration error 0.093).

Figure \ref{fig:redshift_binned_photo_z_mainsample} further shows the model's performance evaluated in redshift bins of $\Delta z=0.5$. Both $\sigma_{\mathrm{NMAD}}$ and $\eta$ remain relatively low at intermediate redshifts, while they increase toward higher redshifts, peaking around $z\sim3.25$. This rise in the outlier fraction at higher redshifts is likely driven by the bias in the training sample, which contains fewer representative sources in this redshift bin and, therefore, limits the ability of the model to predict uniformly across the entire redshift range. Furthermore, at $z\geqslant 3$, \textit{g}-band detections are significantly attenuated by IGM absorption \citep[e.g.,][]{Madau1995, Inoue2014} and detections become increasingly incomplete. While this attenuation produces the strong Lyman break signature that is highly informative for color-based redshift estimation, it also pushes the \textbf{g}-band magnitude closer to the survey detection limit, reducing the signal-to-noise ratio of the ZTF light curves and thereby degrading the quality of the extracted variability features.
The apparent improvement in both metrics at $z>3.5$ should be interpreted with caution, as it is likely driven by the very small number of sources (N = 51) in this bin rather than a genuine recovery in predictive performance. We also note that the redshift-dependent performance trends may be partly influenced by strong emission lines (e.g., Ly$\alpha$, C\,{\sc iv}, Mg\,{\sc ii}) shifting through the observed photometric filters. At specific redshifts, these features can either enhance or degrade the discriminating power of the broadband colors, contributing to the observed trend in the binned metrics.

\begin{figure}[ht!]
    \centering
    \includegraphics[width=1\linewidth]{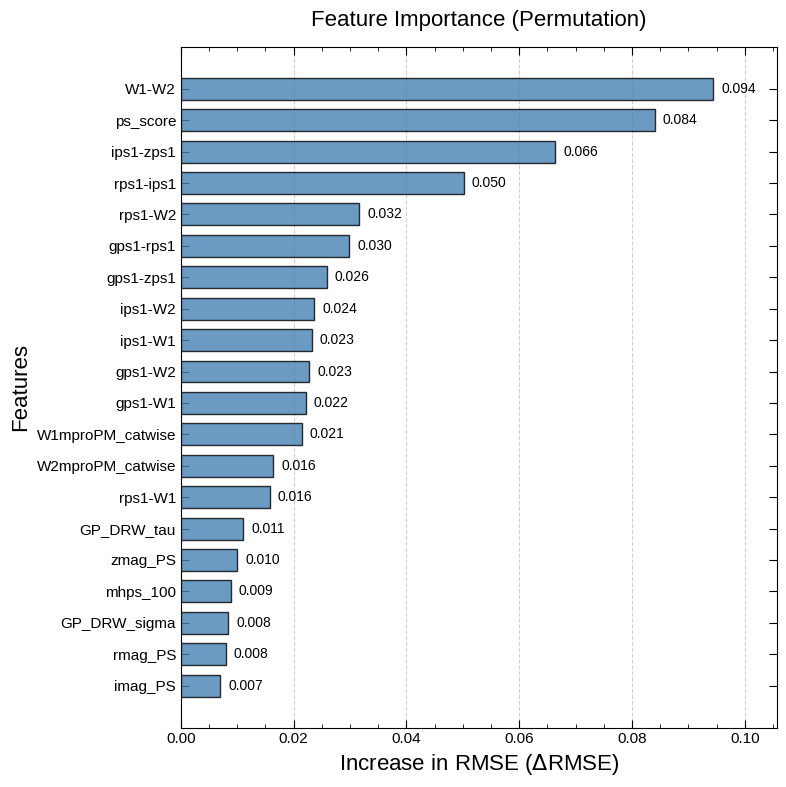}
    \caption{Relative feature importance for the main sample model, determined via permutation importance. The x-axis represents the increase in $\Delta$RMSE when a specific feature is randomly shuffled. Values appended to the bars indicate the exact $\Delta$RMSE for the top 20 features.}
    \label{fig:permutation_importance_main}
\end{figure}

Figure \ref{fig:permutation_importance_main} presents the 20 features that contribute the most to the prediction of AGN photo-$z$s. The complete ranking for the entire feature set is provided in Appendix~\ref{appendix:feature_importance}. The importance is quantified by the increase in the $\Delta$RMSE when a feature is permuted, effectively measuring the model's dependence on that specific feature.

The results show that the MIR color index ($W1-W2$) and the morphology flag from PS1 (\texttt{ps\_score}) are highly ranked, with $\Delta\mathrm{RMSE}$ values of 0.094 and 0.084, respectively. These features are followed by a series of optical and optical-IR color indices, such as \texttt{ips1-zps1}, \texttt{rps1-ips1}, and \texttt{rps1-W2}, which characterize the slope and shape of the AGN SED.

Variability-related features, such as the DRW timescale (\texttt{GP\_DRW\_tau}), the Mexican-hat filter amplitude (\texttt{mhps\_100}), and the DRW amplitude (\texttt{GP\_DRW\_sigma}), are ranked lower in the top 20 compared to photometric and morphological data. This result indicates that while the model utilizes variability as a  secondary constraint, the primary drivers for the redshift estimation in the main sample are the morphology and the broadband color information.

In addition to this analysis, to evaluate the relative importance and contribution of each feature set to the model’s performance, we conducted a systematic ablation study using the same network architecture. We trained and tested the model on four distinct configurations: (i) optical colors combined with variability features (excluding MIR features), (ii) optical colors alone, (iii) variability features alone, and (iv) a combination of optical and MIR colors without variability information. The morphology flag from PS1 was included as a common feature across all four experimental sets. This systematic comparison allows us to quantify the individual predictive power of each data set and assess how different feature combinations impact the accuracy and robustness of the photo-$z$ estimation.

\begin{figure*}[h]
    \centering
    \includegraphics[width=1\linewidth]{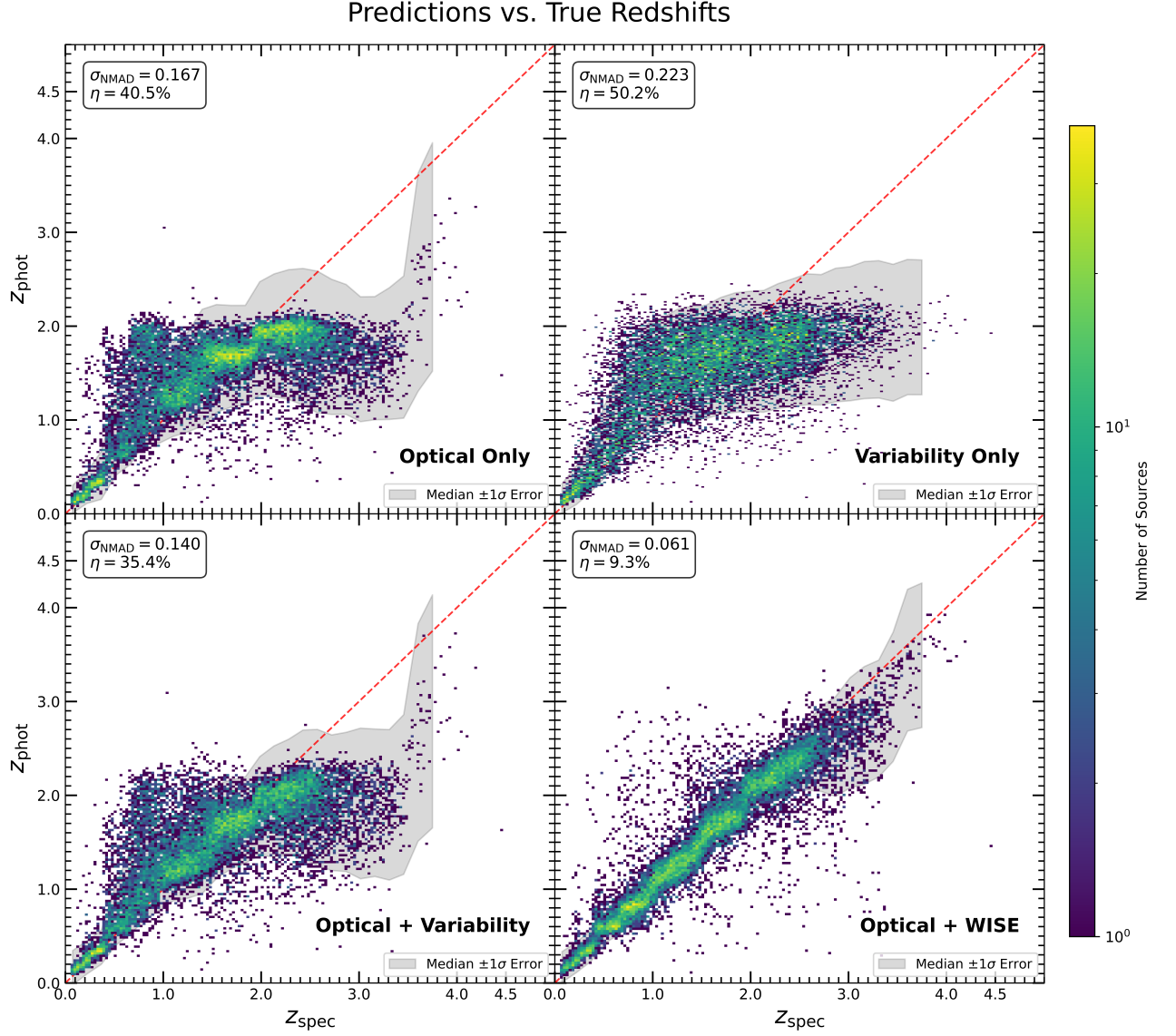}
    \caption{Performance of our model across four distinct feature configurations: optical colors only (top-left), variability features only (top-right), optical colors combined with variability (bottom-left), and optical colors combined with WISE MIR photometry (bottom-right). The red dashed line represents the one-to-one relation. The shaded grey region in each panel indicates the median predicted 
$\pm 1\sigma$ uncertainty.  The corresponding $\sigma_{\mathrm{NMAD}}$ and $\eta$ are mentioned in each panel. For all four configurations, the PS1 morphology flag was included as a constant input feature to the model.}
    \label{fig:results_combinations}
\end{figure*}

\begin{table*}[ht]
\centering
\caption{Results comparing the photo-$z$ performance of different feature sets.}

\label{tab:ablation_results}
\begin{tabular*}{\textwidth}{@{\extracolsep{\fill}}lccc}\toprule
\textbf{Input Features} & \textbf{$\eta$ (\%)} & \textbf{$\sigma_{\text{NMAD}}$(Total sources)} & \textbf{$\sigma_{\text{NMAD}}$(Without outliers)} \\ 
\midrule
Colors (Opt + MIR) + Var & 8.2  & 0.058 & 0.053\\ 
Colors (Opt + MIR)       & 9.3  & 0.061  & 0.054 \\
Colors (Opt) + Var       & 35.4 & 0.140 & 0.079 \\
Optical colors only      & 40.5&0.167 & 0.085 \\
Variability only         & 50.2 &0.223 & 0.105 \\
\bottomrule
\end{tabular*}
\tablefoot{The PS1 morphology flag (\texttt{ps\_score}) is included as a constant input feature in all configurations.}
\end{table*}

The results of our ablation analysis, summarized in Table~\ref{tab:ablation_results} and visualized in Figure~\ref{fig:results_combinations}, highlight the synergy between multi-wavelength photometry and time-domain features for accurate AGN redshift estimation. As shown in the top panels of Figure~\ref{fig:results_combinations}, models using
optical colors alone ($\eta=40.5\%$) or variability features alone ($\eta=50.2\%$) exhibit significant dispersion and a restricted capacity to map the high-redshift regime, where the $griz$ filters no longer effectively capture the redshifted spectral features of the AGNs. While the combination of optical and variability features (bottom-left) provides a measurable improvement by reducing the outlier fraction to 35.4\%, the most substantial performance improvement occurs with the inclusion of MIR data. The optical + MIR configuration (bottom-right) extends the model's reach into the higher redshift regime, producing a significant improvement in accuracy ($\eta=9.3\%$) and highlighting the role of infrared photometry in anchoring the SED and breaking the color-redshift degeneracies that affect the optical-only samples. Ultimately, our full model (Optical + MIR + Variability) achieves the highest overall precision with an outlier fraction of 8.2\%, demonstrating that while MIR data provides the fundamental physical baseline for mapping the entire redshift distribution, the variability features serve as a vital refiner that optimizes the model's reliability across the entire parameter space.

The predicted uncertainty envelopes, shown as shaded grey regions in Figure~\ref{fig:results_combinations}, further illustrate these differences. Configurations with limited wavelength coverage, particularly the optical-only and variability-only models, show wider $\pm 1\sigma$ envelopes at intermediate and high redshifts, reflecting the increased degeneracy in the color-redshift mapping. In contrast, the full model and the optical + MIR configuration display tighter envelopes that closely trace the one-to-one relation, confirming that the MIR information not only reduces the outlier fraction but also significantly improves the precision.

\subsection{Application to UKIDSS subsample}

Using the same model architecture described in Section ~\ref{Method}, we repeated the analysis for the UKIDSS subsample of 47\,906 sources to estimate the limit to which NIR photometry can complement LSST-like optical data for photo-$z$ estimations, when MIR information from WISE is unavailable, which will often be the case for the faint new AGNs LSST will discover, whose fluxes fall below the CatWISE detection limit. This analysis provides a useful proxy for studying how \textit{Euclid} NIR photometry may complement LSST-like optical data in AGN photo-$z$ estimation. 

\begin{figure*}[h]
    \centering
    \includegraphics[width=1\linewidth]{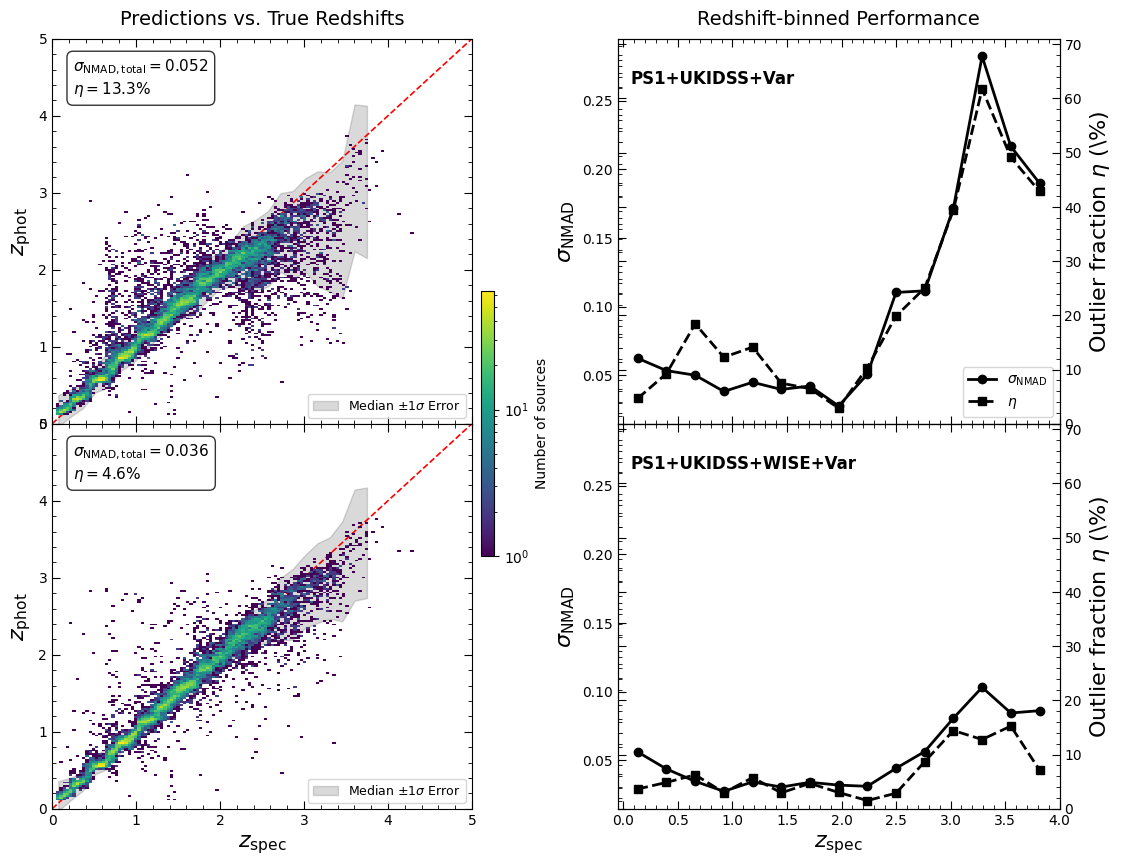}
    \caption{Performance of our photo-$z$ estimator model using AGN variability features combined with PS1+UKIDSS photometry (top row) and PS1+UKIDSS+WISE photometry (bottom row). Left panels: photometric versus spectroscopic redshifts for the test sample, with the color scale indicating the number of sources per bin. The red dashed line represents the one-to-one relation, and the shaded grey region indicates the median predicted $\pm 1\sigma$ uncertainty envelope. The corresponding $\sigma_{\mathrm{NMAD}}$ and outlier fraction ($\eta$) are shown in the inset. Right panels: redshift-binned $\sigma_{\mathrm{NMAD}}$ (solid line with circles) and outlier fraction $\eta$ (dashed line with squares) as a function of spectroscopic redshift.}
    \label{fig:results_subsample}
\end{figure*}

We trained the model on 60\% of the UKIDSS subsample with a dropout rate of $0.2$, validated it on 20\% of the sample, and used the remaining 20\% as an independent test set. Furthermore, for the same UKIDSS subsample, we analyzed an additional configuration in which the PS1, UKIDSS, and WISE photometry were combined. Owing to the larger number of input features in this case, we adopted a higher dropout rate of $0.35$ to mitigate overfitting and to ensure a more stable training of the model.

Figure \ref{fig:results_subsample} compares the photo-$z$ performance of our model for the two photometric combinations, PS1+UKIDSS+variability (top row) and PS1+UKIDSS+WISE+variability (bottom row). The left column shows the comparison between the predicted photo-$z$s and the spec-$z$s. For the PS1+UKIDSS+variability case, we obtain $\sigma_{\mathrm{NMAD}}=0.052$ and an outlier fraction of $\eta=13.3\%$. When the WISE photometry is included, the performance improves significantly to $\sigma_{\mathrm{NMAD}}=0.036$ and $\eta=4.6\%$. This demonstrates that the addition of MIR information provides a clear gain in the overall accuracy of the predictions and reduces the fraction of catastrophic outliers. The predicted $\pm 1\sigma$ uncertainty envelopes are also shown as shaded grey regions in the left panels. For the PS1+UKIDSS+variability configuration, the envelope widens beyond $z \gtrsim 2.5$, reflecting the reduced constraining power of optical and NIR colors at higher redshifts. The inclusion of MIR photometry narrows the envelope across the full redshift range, which is consistent with the marked improvement in the outlier fraction.

The right column of Figure \ref{fig:results_subsample} presents the redshift-binned performance of $\sigma_{\mathrm{NMAD}}$ and $\eta$ for the two cases. The PS1+UKIDSS+variability combination shows a stronger degradation at higher redshift, with both the scatter and the outlier fraction increasing substantially in the highest-$z$ bins. In contrast, the PS1+UKIDSS+WISE+variability model remains much more stable over the full redshift range, with only minor degradation at the highest redshifts. When compared with the main-sample result shown in Figure \ref{fig:photoz_predictions_mainsample}, where we obtained $\eta=8.2\%$ and $\sigma_{\mathrm{NMAD}}=0.058$, we find that although the best overall performance is obtained when WISE photometry is included, the PS1+UKIDSS+variability configuration still yields highly competitive results. This result proves that the combination of optical and NIR data with variability information is effective for AGN photo-$z$ estimation. Consequently, this method is highly promising for forecasting future synergies between LSST, \textit{Euclid}, and \textit{Roman}, where NIR photometry will provide crucial additional constraints on AGN photo-$z$ estimations in the absence of MIR coverage. Furthermore, given that our feature importance analysis (Figure~\ref{fig:permutation_importance_main}) identifies morphology as a highly informative predictor, \textit{Euclid}'s space-based angular resolution is expected to yield even greater improvements by providing highly accurate morphological characterizations of the host galaxies that ground-based surveys cannot match.

\subsection{Comparison with empirical methods}\label{sec:comparison}

To benchmark our model against established physical and empirical approaches, we compared its performance with SED modeling and variability modeling photo-$z$ frameworks. This comparison was performed on the test sample of 9605 sources with optical+NIR+MIR photometric data coverage from PS1, UKIDSS, and CatWISE. For the SED modeling component, we performed template-fitting using Low-Resolution Templates \citep[LRT;][]{Assef2010}. Simultaneously, we implemented the \texttt{VAR-PZ} framework, utilizing \texttt{LRT} as its underlying photometric baseline. To the SED results, we applied the \texttt{VAR-PZ} variability priors derived from ZTF \textit{g}-band light curves \citepalias{arevalo2025}, which covers an average baseline of 1400 days (including seasonal gaps). Following the baseline requirements mentioned in \citet{VARPZ} and \citet{Suberlak_2021}, we used a $3\tau$ constraint, requiring that the total light curve baseline be at least three times the characteristic damping timescale ($\tau$). This threshold ensures the light curve baseline is sufficient to reliably estimate the DRW parameters while avoiding degeneracies in the model fitting although some authors suggest an even stronger constraints of $10\tau$ \citep[e.g.,][]{Koz_owski_2017}. Consequently, \texttt{VAR-PZ} priors are only applied in redshift regimes where the temporal baseline is sufficient to provide a formal constraint on the variability timescale. Outside of this regime, the total constraining power is given solely to the \texttt{LRT} SED photo-$z$ priors.

\begin{figure}
    \centering
    \includegraphics[width=1\linewidth]{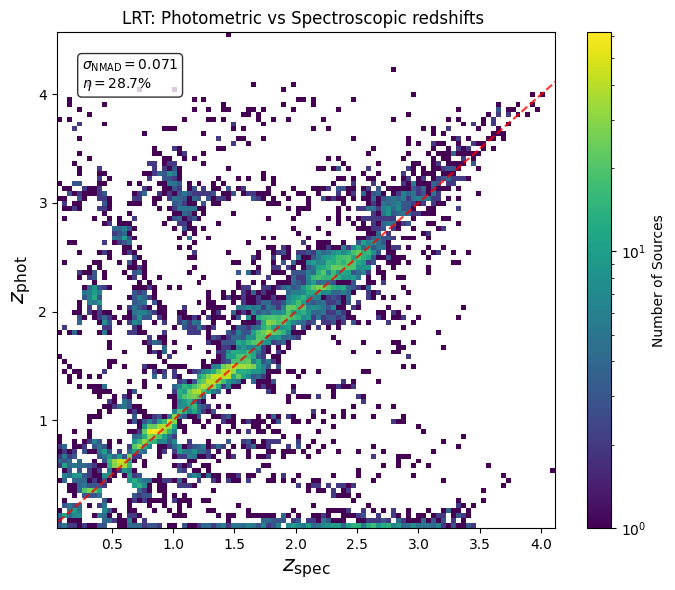}
    \caption{Photometric versus spectroscopic redshifts for the test sample estimated via \texttt{LRT} template-fitting using PS1, UKIDSS, and CatWISE photometry. The red dashed line represents the one-to-one correspondence. The calculated $\sigma_{\mathrm{NMAD}}$ and outlier fraction ($\eta$) are shown in the inset.}
    \label{fig:LRT}
\end{figure}
\begin{figure}
    \centering
    \includegraphics[width=1\linewidth]{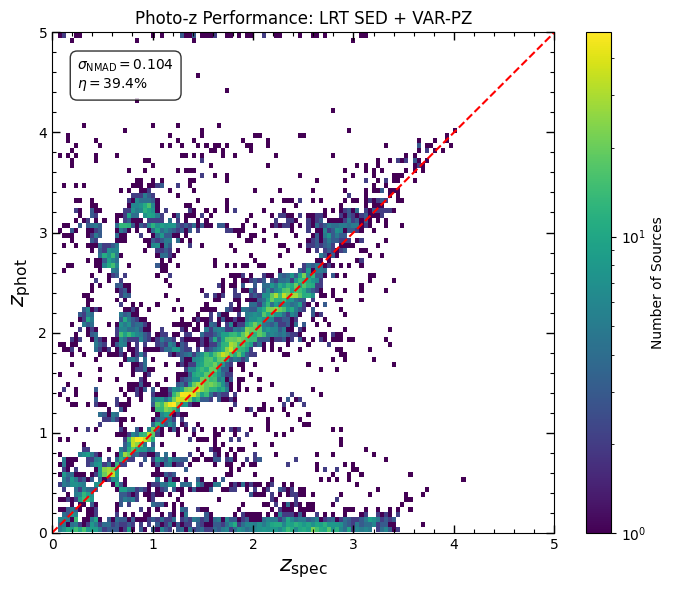}
    \caption{Results for the \texttt{VAR-PZ} framework, illustrating the performance when variability-based constraints from ZTF light curves are combined with \texttt{LRT} SED modeling predictions. The scatter plot compares the resulting photo-$z$ estimates against spectroscopic values for the test sample, with the one-to-one relation indicated by the red dashed line. The inset provides the corresponding statistical metrics, including the outlier fraction ($\eta$) and $\sigma_{\mathrm{NMAD}}$.}
    \label{fig:varpz}
\end{figure}

The comparison of the benchmark results is presented in Figures \ref{fig:LRT} and \ref{fig:varpz}. \texttt{LRT} template-fitting yields an outlier fraction of $\eta=28.7\%$ with $\sigma_{\mathrm{NMAD}}=0.071$. When the \texttt{VAR-PZ} variability constraints are applied on top of the \texttt{LRT} SED priors, the outlier fraction increases to $\eta=39.4\%$ ($\sigma_{\mathrm{NMAD}}=0.104$), indicating that the addition of single-band variability information actively degrades the photo-$z$ performance. This worsening can be attributed to the fact that the \texttt{VAR-PZ} framework was designed to stabilize the correct redshift from the combination of variability information across multiple photometric bands. When applied to a single \textit{g}-band light curve, the DRW model fit yields degenerate solutions that fail to uniquely constrain the redshift, thereby introducing incorrect priors that pull the SED-based estimates away from their true values.

To verify this interpretation, we performed a simulation test using 8237 objects from our testing sample with black hole masses ($M_{\mathrm{BH}}$) available from SDSS \citep{Shen2011}. We simulated ZTF-like \textit{g}-band DRW light curves as described in appendix \ref{appendix:simulations}, and applied the \texttt{VAR-PZ} framework to these simulated light curves. The results show that the outlier fraction increases from $\eta=27.6\%$ (\texttt{LRT} only) to $\eta=28.1\%$ (\texttt{LRT}+\texttt{VAR-PZ}). The worsening in the simulated case is substantially smaller than for the real data because the simulated DRW light curves, by construction, have well-behaved and consistent photometric uncertainties, unlike the real ZTF data which are affected by systematics and non-DRW variability components. The full details of this simulation analysis are presented in Appendix~\ref{appendix:simulations}.

The substantial gain of our ML model over \texttt{LRT} ($\eta=28.7\%$) is primarily driven by the neural network architecture's ability to learn non linear mappings in the multi-wavelength photometric space. By more effectively utilizing the infrared colors to anchor the SED, the model maintains higher reliability across the entire redshift range and avoids the catastrophic failures that often affect traditional template-fitting methods at higher redshifts.

\subsection{Limitations and caveats}\label{caveats}
We note that our labeled sample of 72\,728 sources is drawn from SDSS DR16/DR19 and DESI DR1 spectroscopic catalogs, both of which are subject to well known target selection biases. These surveys preferentially observe optically bright, unobscured (type 1) AGNs, and their targeting strategy is based on color and morphology pre-selections that may not be representative of the full AGN population. As a result, our training sample is likely incomplete for faint or high-$z$ AGNs, precisely the population that LSST is expected to observe in large numbers for the first time. 

Furthermore, our sample is inherently conditioned on the light-curve classification. By restricting our analysis to sources explicitly classified as \texttt{AGN} or \texttt{QSO} based on their ZTF $g$-band variability, we introduce an additional selection bias: the model is trained exclusively on AGNs that are sufficiently variable in the optical to be detected by the random forest classifier. While this is representative of how LSST will identify new AGN candidates from time-domain data, it means our results may not fully generalize to populations of active galaxies with low-amplitude or long-timescale variability that evade initial time-domain selection. 

This selection bias also has implications for the interpretation of the feature importance analysis. The \texttt{ps\_score} morphology flag, which ranks among the most important features, may partly encode properties of the spectroscopic training sample itself rather than purely physical redshift information. In a magnitude-limited spectroscopic survey, brighter and more point-like AGNs are preferentially observed at higher redshifts, introducing a Malmquist-type selection effect. Consequently, the model may learn to associate point-like morphology with higher redshift in a way that reflects the selection function of the training data rather than a universal physical correlation. This effect would limit the model's ability to generalize to sources such as small, lower-redshift host galaxies that can appear point-like, where the learned relationship between morphology and redshift may break down.

The practical consequence is that \texttt{VAR-PZnn} may produce a higher fraction of catastrophic outliers when applied to sources that lie outside the parameter space covered by the current training data. In particular, type 2 AGNs, broad absorption line (BAL) quasars, heavily host-dominated sources, and AGNs at $z > 3.5$ are underrepresented in our sample, and the model's reliability for these populations cannot be guaranteed from the present analysis. When applied to the broader LSST AGN population, which will include significantly fainter and more obscured sources than are represented in our training set, the performance reported here should be regarded as an optimistic estimate. Mitigating this limitation will require augmenting the training set with deeper spectroscopic surveys such as 4MOST and MOONS as they become available, as well as exploring semi-supervised or transfer learning approaches to extend model generalization beyond the current training distribution.

Finally, we note that individual measurement uncertainties on the input features (photometric errors, variability descriptor uncertainties) are not explicitly provided as inputs to the network. This is consistent with the approach adopted by most current ML-based photo-$z$ methods in the literature. However, we note that our MC dropout framework (Section~\ref{subsec:uncertainty}) partially mitigates this limitation by providing a model-level estimate of predictive uncertainty, but this does not replace the information carried by the individual feature errors. Explicitly propagating measurement uncertainties through the network represents a natural extension of this work.

\section{Summary and conclusions} \label{conclusions}

We investigated the integration of AGN variability, multi-band photometric colors, and morphological parameters to estimate AGN photo-$z$s using a fully connected mixture density network architecture. Our analysis was based on 72\,728 \texttt{AGN/QSO} sources with spectroscopic redshifts from SDSS DR19 and DESI DR1, alongside a subsample of 47\,906 sources with additional UKIDSS NIR photometry. The input feature space combined variability features derived from ZTF \textit{g}-band light curves with optical colors from PS1, MIR colors from WISE, and, when available, NIR colors from UKIDSS. The model was trained and evaluated using a 60:20:20 split into training, validation, and test samples. Its performance was assessed through the NMAD ($\sigma_{\mathrm{NMAD}}$), and the outlier fraction ($\eta$). 

For the main sample, which combines variability information with PS1 and WISE photometry, we obtained an overall performance of $\sigma_{\mathrm{NMAD}} = 0.058$ and $\eta = 8.2\%$. The comparison between photometric and spectroscopic redshifts, as shown in Figure \ref{fig:photoz_predictions_mainsample}, illustrates that the model is able to recover the general redshift distribution of our test sample with good accuracy over most of the explored range. In addition to point estimates, our MDN architecture combined with MC 
dropout inference provides per-source photo-$z$ uncertainties 
($\sigma_z$) that decompose into aleatoric and epistemic components. 
The predicted uncertainties are well correlated with the actual 
prediction errors, and the model correctly assigns larger 
uncertainties to sources at higher redshifts where training data is 
sparse. By excluding the 10\% of sources with the highest predicted 
uncertainty, the outlier fraction for the main sample is reduced from 
8.2\% to 5.4\%, demonstrating that the uncertainty estimates provide 
a practical quality flag for catalog-level applications. The redshift-binned analysis indicates that the performance is more stable at low and intermediate redshift, while both the scatter and the outlier fraction increase toward the highest-$z$ bins. This behavior is likely related to the smaller number of representative training sources at high redshift.

We explored the role of NIR photometry using the UKIDSS subsample. For the configuration based on PS1 + UKIDSS, we obtained $\sigma_{\mathrm{NMAD}} = 0.052$ and $\eta = 13.3\%$. When WISE photometry was included, the performance improved to $\sigma_{\mathrm{NMAD}} = 0.036$ and $\eta = 4.6\%$. These results show that the MIR information provides the strongest improvement in the overall photo-$z$ accuracy, but also that the combination of optical, variability, and NIR information alone already yields competitive results.

Our results therefore suggest that NIR photometry can provide useful additional constraints for AGN photo-$z$ estimation, especially in cases where MIR data are unavailable. Consequently, the UKIDSS-based analysis provides an interesting forecast for what can be achieved with \textit{Euclid} and \textit{Roman}, provided that one accounts for the intrinsic differences between the two surveys. In particular, UKIDSS includes the $K$ band, whose colors contribute measurably to the NIR configuration (although $H$- and $Y$-band colors rank highest; see Appendix~\ref{appendix:feature_importance}) and which will be absent from the \textit{Euclid} NISP filter set" ($Y_E$, $J_E$, $H_E$) and the surveys differ significantly in their survey depth, angular resolution, filter transmission profiles (with \textit{Euclid} filters being broader and gapless), and source selection functions. Therefore, the UKIDSS-based analysis should be viewed as an exploratory first estimate rather than a direct forecast of \textit{Euclid} performance. Crucially, because morphology ranks among the most critical features for accurate predictions, \textit{Euclid}'s space-based high-resolution imaging will provide a far superior characterization of host-galaxy morphology compared to the seeing-limited UKIDSS data, which should significantly enhance photo-$z$ performance.

Furthermore, our results represent a significant improvement over established physical and empirical benchmarks. On the same sources with full optical+NIR+MIR coverage, our model (PS1+UKIDSS+WISE+variability, $\eta=4.6\%$, $\sigma{\mathrm{NMAD}}=0.036$) substantially outperforms traditional \texttt{LRT} template-fitting ($\eta=28.7\%$, $\sigma_{\mathrm{NMAD}}=0.071$) evaluated on the identical set. 
Interestingly, directly applying single-band \texttt{VAR-PZ} variability priors to the \texttt{LRT} outputs results in an outlier fraction of $\eta=39.4\%$. Rather than a limitation of the \texttt{VAR-PZ} framework, this behavior highlights the fundamental challenge of degenerate DRW solutions inherent to observed single-band light curves. This physical interpretation is supported by our DRW simulation analysis (Appendix~\ref{appendix:simulations}), which confirms that when \texttt{VAR-PZ} is applied to idealized DRW single-band light curves, the outlier fraction remains stable, increasing marginally from $\eta=27.6\%$ to $28.1\%$. 

This comparison demonstrates that directly learning the non-linear mapping between multi-wavelength photometry and redshift, rather than relying on template matching, effectively mitigates the color-redshift degeneracies that limit traditional SED-fitting approaches. Ultimately, the dominant gain in AGN photo-$z$ performance arises from MIR photometry and morphological constraints, while single-band variability provides a secondary but measurable improvement when incorporated into a probabilistic multi-wavelength framework. Looking ahead, the combination of multi-band variability from LSST, along with the addition of the LSST $u$-band and the much higher spatial resolution and deeper NIR photometry from \textit{Euclid},  as well as deep NIR imaging from the Nancy Grace Roman Space Telescope, will provide a significantly richer feature space. These improvements, which are not captured in our current analysis, are expected to further mitigate color-redshift degeneracies and improve both the precision and completeness of the photometric redshift catalogs required by the next generation of large-scale AGN studies. Finally, we note that our analysis is based on ZTF \textit{g}-band light curves spanning approximately 4.5 years (up to September 2022). The ZTF survey has now accumulated over 8 years of observations, which would improve the reliability of the DRW parameter estimation, particularly for sources with longer characteristic timescales. As the LSST accumulates data over its 10-year survey, both the number of sampled variability timescales and the photometric precision of the extracted features will steadily increase.

\section*{Data and code availability}
The code and trained models used in this study are publicly available in the GitHub repository at \url{https://github.com/SarathSS98/varpznn}. For further details regarding the dataset compilation, we refer the reader to \citet{arevalo2025}.

\begin{acknowledgements}

 We gratefully acknowledge the support of the ANID BASAL project FB210003 (S.S.S., R.J.A., T.A., F.E.B., M.E.O., C.M., T.M.), FONDECYT Regular 1231718 (S.S.S. and R.J.A.), 1240105 (T.A. and M.E.O.), 1241422 (P.A.), FONDECYT iniciaci\'on grant 11240336 (C.M. and T.M) and 11241477 (L.H.G.). T.T.A acknowledges support from NASA ADAP Grant 80NSSC24K0692. DD, VP and MF acknowledge the financial support by Italian Ministry for Education University and Research (MIUR) grant PRIN 2022 2022383WFT “SUNRISE” and from the Timedomes grant within the “INAF 2023 Finanziamento della Ricerca Fondamentale”. Support was provided by Schmidt Sciences, LLC. for J.F. A.B.K. and D.I. acknowledge funding provided by the University of Belgrade - Faculty of Mathematics (the contract 451-03-136/2025-03/200104) through the grants by the Ministry of Science, Technological Development and Innovation of the Republic of Serbia. RAR  acknowledges the support from the Conselho Nacional de Desenvolvimento Científico e Tecnológico (CNPq; Project 303450/2022-3) and Fundação de Amparo à Pesquisa do Estado do Rio Grande do Sul (FAPERGS; Project 25/2551-0002765-9). C.G.B. acknowledges support from the Consejo Nacional de Investigaciones Científicas y Técnicas (CONICET) and the Secretaría de Ciencia y Tecnología de la Universidad Nacional de Córdoba (SeCyT). D.M. acknowledges financial support from CAPES – Finance Code 001. W.N.B. acknowledges the support of USA NSF grant AST-2407089. SP is supported by the international Gemini Observatory, a program of NSF NOIRLab, which is managed by the Association of Universities for Research in Astronomy (AURA) under a cooperative agreement with the U.S. National Science Foundation, on behalf of the Gemini partnership of Argentina, Brazil, Canada, Chile, the Republic of Korea, and the United States of America.  M.J.T. acknowledges funding from UKRI grant ST/X001075/1.

\newline

\end{acknowledgements}

 \textit{Software packages}: \texttt{PyTorch} \citep{PyTorch}, \texttt{scikit-learn} \citep{scikit-learn}, \texttt{NumPy} \citep{2020Numpy}, \texttt{Astropy} \citep{2018Astropy}, \texttt{Matplotlib} \citep{Matplotlib}, \texttt{Pandas} \citep{reback2020pandas}.

\bibliographystyle{aa}
\bibliography{ref}

\onecolumn
\appendix

\section{Description of input features}

Table \ref{appendix:features_table} presents a comprehensive list of all the input features provided to the \texttt{VAR-PZnn} model. These features are grouped into four broad categories: variability metrics derived from ZTF light curves, broadband photometry from optical and IR surveys, photometric colors, and host-galaxy morphology indicator. The description and original reference for each feature are also provided.
\begin{table}[h]
\centering
\caption{Detailed summary of the input features used in our analysis, categorized into variability metrics, broadband photometry, photometric colors, and morphological indicators.}
\label{appendix:features_table}
\small
\setlength{\tabcolsep}{5pt}
\begin{tabularx}{\textwidth}{l X l}
\toprule
\textbf{Feature} & \textbf{Description} & \textbf{Reference} \\
\midrule

\multicolumn{3}{l}{\textbf{Variability features}} \\
\midrule
Pvar & Probability that the source is intrinsically variable & \citetalias{alerts_classifier,arevalo2025} \\
ExcessVar & Measure of the intrinsic variability amplitude & \citetalias{alerts_classifier,arevalo2025} \\
MHPS parameters$^\ast$ & Mexican-Hat analysis at different timescales (6 features in total) & \citetalias{alerts_classifier,arevalo2025} \\
Anderson Darling & Test of whether a sample of data comes from a population with a specific distribution & \citetalias{alerts_classifier,arevalo2025} \\
Autocor\_length & Lag value where the autocorrelation function becomes smaller than Eta\_e & \citetalias{alerts_classifier,arevalo2025} \\
Con & Number of three consecutive data points brighter/fainter than $2\sigma$ of the light curve & \citetalias{alerts_classifier,arevalo2025} \\
Eta\_e & Ratio of the mean of the squares of successive magnitude differences to the variance of the light curve & \citetalias{alerts_classifier,arevalo2025} \\
Gskew & Median-based measure of the skewness & \citetalias{alerts_classifier,arevalo2025} \\
Mean & Mean magnitude of the light curve & \citetalias{arevalo2025} \\
Mean variance & Ratio of the standard deviation to the mean magnitude & \citetalias{alerts_classifier,arevalo2025} \\
Q31 & Difference between the third and first quartiles of the light curve & \citetalias{alerts_classifier,arevalo2025} \\
Rcs & Range of a cumulative sum & \citetalias{alerts_classifier,arevalo2025} \\
Std & Standard deviation of the light curve & \citetalias{alerts_classifier,arevalo2025} \\
StetsonK & Robust kurtosis measure & \citetalias{alerts_classifier,arevalo2025} \\
SF\_ML\_amplitude & RMS magnitude difference of the structure function, computed over a 1 yr timescale & \citetalias{alerts_classifier,arevalo2025} \\
SF\_ML\_gamma & Logarithmic gradient of the mean change in magnitude & \citetalias{alerts_classifier,arevalo2025} \\
IAR\_phi & Level of autocorrelation using a discrete-time representation of a DRW model & \citetalias{alerts_classifier,arevalo2025} \\
GP\_DRW\_sigma & Amplitude of the variability at short timescales ($t=\tau$), from DRW modeling & \citetalias{alerts_classifier,arevalo2025} \\
GP\_DRW\_tau & Relaxation time $\tau$ from DRW modeling & \citetalias{alerts_classifier,arevalo2025} \\
Psi\_CS\_v2 & Range of a cumulative sum applied to the phase-folded light curve & \citetalias{alerts_classifier,arevalo2025} \\
Psi\_eta\_v2 & Eta\_e index calculated from the folded light curve & \citetalias{alerts_classifier,arevalo2025} \\

\midrule
\multicolumn{3}{l}{\textbf{Photometry features}} \\
\midrule
gmag\_PS & \textit{g}-band magnitude from PS1 & \citetalias{sanchez_ZTF_2023,arevalo2025} \\
rmag\_PS & $r$-band magnitude from PS1 & \citetalias{sanchez_ZTF_2023,arevalo2025} \\
imag\_PS & $i$-band magnitude from PS1 & \citetalias{sanchez_ZTF_2023,arevalo2025} \\
zmag\_PS & $z$-band magnitude from PS1 & \citetalias{sanchez_ZTF_2023,arevalo2025} \\
Y & $Y$-band magnitude from UKIDSS & This work \\
J & $J$-band magnitude from UKIDSS & This work \\
H & $H$-band magnitude from UKIDSS & This work \\
K & $K$-band magnitude from UKIDSS & This work \\
W1 & W1-band magnitude from CatWISE & \citetalias{sanchez_ZTF_2023,arevalo2025} \\
W2 & W2-band magnitude from CatWISE & \citetalias{sanchez_ZTF_2023,arevalo2025} \\

\midrule
\multicolumn{3}{l}{\textbf{Color features}} \\
\midrule

Optical colors$^\ast$ & Colors derived from the PS1 \textit{griz} bands & \citetalias{sanchez_ZTF_2023,arevalo2025} \\
Optical-NIR colors$^\ast$ & Colors combining PS1 \textit{griz} and UKIDSS $YJHK$ photometry (16 in total) & This work \\
NIR colors$^\ast$ & Colors derived from the UKIDSS $YJHK$ bands (6 in total) & This work \\
Optical-MIR colors$^\ast$ & Colors combining PS1 optical and WISE MIR photometry (6 in total) & \citetalias{sanchez_ZTF_2023} \\
NIR-MIR colors$^\ast$ & Colors combining UKIDSS $YJHK$ and WISE $W1,W2$ photometry & This work \\
MIR colors$^\ast$ & Color derived from the WISE W1 \& W2 bands & \citetalias{sanchez_ZTF_2023,arevalo2025} \\
\midrule
\multicolumn{3}{l}{\textbf{Morphology feature}} \\
\midrule
ps\_score &  Morphology indicator from PS1  & \citetalias{sanchez_ZTF_2023,arevalo2025} \\

\bottomrule
\label{Features_table}
\end{tabularx}
\tablefoot{$^\ast$Features marked with an asterisk denote groups of related quantities. MHPS parameters include \texttt{mhps\_10}, \texttt{mhps\_45}, \texttt{mhps\_100}, \texttt{mhps\_450}, \texttt{ratio\_mhps\_100\_10}, and \texttt{ratio\_mhps\_450\_45}. Color features correspond to the subset of pairwise magnitude combinations adopted in this work; the complete list is given in Appendix~\ref{appendix:feature_importance}.}
\end{table}

\newpage
\section{Neural Network Workflow}
\begin{figure*}[h]
    \centering
    \includegraphics[width=1\linewidth]{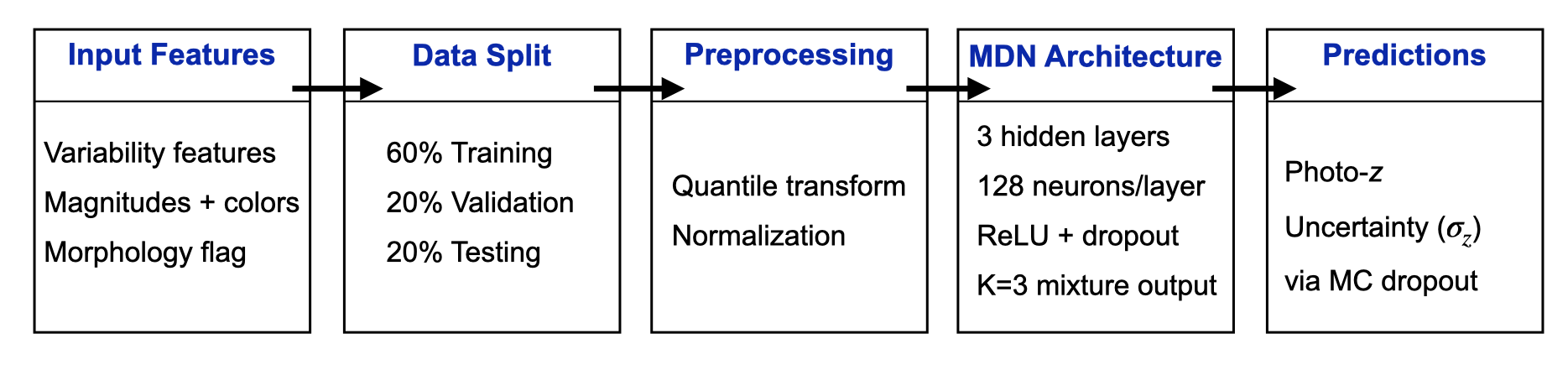}
    \caption{Workflow of our fully connected mixture density network model used for AGN photo-$z$ estimation.}
    \label{fig:flowchart}
\end{figure*}

Figure \ref{fig:flowchart} provides a schematic overview of the \texttt{VAR-PZnn} architecture. The framework takes in optical and infrared photometry, variability parameters, and a morphology flag. These inputs are combined and processed by three hidden layers (with 128 neurons each, using LeakyReLU activations and dropout). Instead of predicting a single redshift value, the final MDN layer outputs a probability distribution. Specifically, it predicts the weights ($\pi_k$), means ($\mu_k$), and standard deviations ($\sigma_k$) of three Gaussian components ($K=3$) to model the full photo-$z$ posterior probability $p(z \mid \mathbf{x})$. This probabilistic approach helps capture the complex relationships and degeneracies in the color-redshift mapping.

\label{appendix:workflow}

\section{Network hyperparameters: Gaussian Mixture Components}
\label{appendix:Gaussian}
To determine the optimal number of Gaussian components ($K$) for our Mixture Density Network, we conducted a sensitivity test evaluating the network's performance across $K \in \{1, 2, 3, 4, 5\}$. The results, obtained using the main sample configuration, are summarized in Table~\ref{tab:k_comparison}.
We find that the model's performance is robust to the choice of $K$, with the outlier fraction varying by at most 0.5 percentage points across all configurations. This stability indicates that the photo-$z$ posteriors for the majority of the sources in our sample are well-characterized even by a single Gaussian, and the additional components primarily improve the fit for the minority of sources with genuinely multi-modal posteriors. We adopt $K=3$ as it provides sufficient flexibility to capture potential multi-modality without introducing unnecessary model complexity.
\begin{table}[H]
    \centering
    \caption{Performance comparison of the MDN architecture on the test sample as a function of the number of Gaussian mixture components ($K$).}
    \begin{tabular}{ccc}
        \hline\hline
        $K$ components & $\sigma_{\mathrm{NMAD}}$ & $\eta$ (\%) \\
        \hline
        1 & 0.0619 & 8.7 \\
        2 & 0.0616 & 8.5 \\
        3 & 0.0582 & 8.2 \\
        4 & 0.0589 & 8.2 \\
        5 & 0.0593 & 8.3 \\
        \hline
    \end{tabular}
    \label{tab:k_comparison}
\end{table}

\section{Feature importance}
\label{appendix:feature_importance}

The complete results of our permutation feature importance analysis in Table \ref{appendix:feature_importance}. The importance of each feature is quantified by the increase in the $\Delta\mathrm{RMSE}$ when the values of that feature are randomly shuffled across the test sample. By breaking the true association between the feature and the target redshift while preserving its marginal distribution, this metric isolates the predictive power of each input. The table summarizes the 
$\Delta\mathrm{RMSE}$ scores for all variability features, multi-wavelength colors, and the morphology flag across the distinct ablation scenarios discussed in Section \ref{results}. The top five most influential features for each specific model configuration are highlighted in bold to illustrate how the network shifts its reliance on different physical tracers depending on the available multi-wavelength coverage.

\setlength{\LTleft}{0pt}
\setlength{\LTright}{3pt}

\begin{longtable}{@{\extracolsep{\fill}} l c c c c c c c }
\caption{Full feature importance ($\Delta$RMSE) for all ablation studies. Bolds indicate the top 5 most important features for that specific study.} \label{tab:feat_imp} \\
\toprule
\textbf{Feature} & \makecell[b]{\textbf{Var+Opt}\\\textbf{+MIR}} & \textbf{Var+Opt} & \textbf{Opt Colors} & \makecell[b]{\textbf{Opt}\\\textbf{+MIR}} & \textbf{Var} & \makecell[b]{\textbf{Var+Opt}\\\textbf{+NIR+MIR}} & \makecell[b]{\textbf{Var+Opt}\\\textbf{+NIR}} \\
\midrule
\endfirsthead
\toprule

\midrule
\endhead
\midrule \multicolumn{8}{r}{Continued on next page} \\
\bottomrule
\endfoot
\bottomrule
\endlastfoot
\midrule \multicolumn{8}{l}{\textbf{Morphology}} \\ \midrule
ps\_score & \textbf{0.0840} & \textbf{0.1481} & \textbf{0.1787} & \textbf{0.1065} & \textbf{0.1522} & \textbf{0.1456} & \textbf{0.1187} \\
\midrule \multicolumn{8}{l}{\textbf{Variability}} \\ \midrule
GP\_DRW\_tau & 0.0111 & \textbf{0.0386} & --- & --- & \textbf{0.0388} & 0.0058 & 0.0161 \\
GP\_DRW\_sigma & 0.0084 & \textbf{0.0379} & --- & --- & \textbf{0.0434} & 0.0039 & 0.0118 \\
ExcessVar & 0.0056 & 0.0038 & --- & --- & 0.0068 & 0.0063 & 0.0082 \\
Std & 0.0044 & 0.0030 & --- & --- & 0.0128 & 0.0007 & 0.0061 \\
Mean & 0.0040 & \textbf{0.0562} & --- & --- & \textbf{0.0980} & 0.0064 & 0.0139 \\
Eta\_e & 0.0068 & 0.0086 & --- & --- & 0.0092 & 0.0037 & 0.0066 \\
SF\_ML\_amplitude & 0.0025 & 0.0044 & --- & --- & 0.0045 & 0.0020 & 0.0031 \\
Rcs & 0.0009 & 0.0012 & --- & --- & 0.0006 & 0.0001 & 0.0021 \\
Q31 & 0.0015 & 0.0010 & --- & --- & 0.0029 & 0.0015 & 0.0032 \\
Psi\_eta\_v2 & 0.0005 & 0.0008 & --- & --- & 0.0026 & 0.0010 & -0.0003 \\
Pvar & 0.0010 & 0.0003 & --- & --- & 0.0028 & 0.0016 & 0.0019 \\
Meanvariance & 0.0036 & 0.0029 & --- & --- & 0.0074 & 0.0017 & 0.0011 \\
Psi\_CS\_v2 & 0.0003 & 0.0006 & --- & --- & 0.0009 & 0.0002 & 0.0006 \\
AndersonDarling & 0.0002 & 0.0004 & --- & --- & 0.0023 & -0.0004 & 0.0004 \\
SF\_ML\_gamma & 0.0006 & 0.0019 & --- & --- & 0.0014 & 0.0005 & 0.0011 \\
Con & 0.0001 & 0.0002 & --- & --- & 0.0000 & 0.0011 & 0.0005 \\
mhps\_100 & 0.0089 & 0.0112 & --- & --- & \textbf{0.0175} & 0.0082 & 0.0094 \\
mhps\_450 & 0.0038 & 0.0015 & --- & --- & 0.0059 & 0.0039 & 0.0017 \\
mhps\_45 & 0.0028 & 0.0043 & --- & --- & 0.0054 & 0.0022 & 0.0030 \\
ratio\_mhps\_100\_10 & 0.0019 & 0.0095 & --- & --- & 0.0106 & 0.0002 & 0.0063 \\
ratio\_mhps\_450\_45 & 0.0006 & 0.0018 & --- & --- & 0.0044 & 0.0003 & 0.0026 \\
IAR\_phi & 0.0008 & 0.0025 & --- & --- & 0.0031 & 0.0002 & 0.0015 \\
Autocor\_length & 0.0012 & 0.0024 & --- & --- & 0.0034 & 0.0019 & 0.0034 \\
StetsonK & 0.0005 & 0.0000 & --- & --- & 0.0002 & 0.0006 & 0.0005 \\
mhps\_10 & 0.0032 & 0.0034 & --- & --- & 0.0097 & 0.0030 & 0.0021 \\
Gskew & 0.0012 & -0.0003 & --- & --- & 0.0001 & 0.0008 & -0.0004 \\
\midrule \multicolumn{8}{l}{\textbf{Photometry \& Color}} \\ \midrule
gmag\_PS & 0.0063 & 0.0079 & 0.0148 & 0.0102 & --- & 0.0006 & 0.0047 \\
rmag\_PS & 0.0079 & 0.0056 & \textbf{0.0377} & 0.0123 & --- & 0.0044 & 0.0068 \\
imag\_PS & 0.0070 & 0.0061 & 0.0238 & 0.0145 & --- & 0.0034 & 0.0089 \\
zmag\_PS & 0.0100 & 0.0066 & 0.0148 & 0.0120 & --- & 0.0053 & 0.0075 \\
hapermag3 & --- & --- & --- & --- & --- & 0.0056 & 0.0028 \\
yapermag3 & --- & --- & --- & --- & --- & 0.0022 & 0.0041 \\
kapermag3 & --- & --- & --- & --- & --- & 0.0038 & 0.0033 \\
j\_1apermag3 & --- & --- & --- & --- & --- & 0.0027 & 0.0098 \\

W1mproPM\_CatWISE & 0.0214 & --- & --- & 0.0374 & --- & 0.0044 & --- \\
W2mproPM\_CatWISE & 0.0164 & --- & --- & 0.0080 & --- & 0.0030 & --- \\

W1-W2 & \textbf{0.0944} & --- & --- & \textbf{0.0839} & --- & \textbf{0.0336} & --- \\
ips1-zps1 & \textbf{0.0664} & \textbf{0.0340} & \textbf{0.0388} & \textbf{0.0676} & --- & \textbf{0.0190} & \textbf{0.0485} \\
ips1-W2 & 0.0237 & --- & --- & 0.0440 & --- & 0.0110 & --- \\
rps1-ips1 & \textbf{0.0502} & 0.0322 & \textbf{0.0292} & \textbf{0.0527} & --- & \textbf{0.0163} & 0.0197 \\
ips1-W1 & 0.0232 & --- & --- & 0.0296 & --- & 0.0024 & --- \\
gps1-rps1 & 0.0299 & 0.0275 & \textbf{0.0276} & 0.0277 & --- & \textbf{0.0250} & \textbf{0.0394} \\
gps1-zps1 & 0.0258 & 0.0133 & 0.0258 & 0.0249 & --- & 0.0062 & 0.0067 \\
rps1-W2 & \textbf{0.0316} & --- & --- & \textbf{0.0625} & --- & 0.0082 & --- \\
gps1-W2 & 0.0228 & --- & --- & 0.0453 & --- & 0.0057 & --- \\
rps1-W1 & 0.0158 & --- & --- & 0.0259 & --- & 0.0023 & --- \\
gps1-W1 & 0.0221 & --- & --- & 0.0114 & --- & 0.0046 & --- \\

Hmag-W1 & --- & --- & --- & --- & --- & 0.0053 & --- \\
Ymag-W1 & --- & --- & --- & --- & --- & 0.0048 & --- \\
Kmag-W1 & --- & --- & --- & --- & --- & 0.0114 & --- \\
Ymag-Hmag & --- & --- & --- & --- & --- & 0.0064 & 0.0226 \\
Hmag-Kmag & --- & --- & --- & --- & --- & 0.0032 & 0.0256 \\
zmag\_PS-Hmag & --- & --- & --- & --- & --- & 0.0051 & 0.0127 \\
Jmag-W1 & --- & --- & --- & --- & --- & 0.0046 & --- \\
Hmag-W2 & --- & --- & --- & --- & --- & 0.0106 & --- \\
Jmag-W2 & --- & --- & --- & --- & --- & 0.0074 & --- \\
Kmag-W2 & --- & --- & --- & --- & --- & 0.0136 & --- \\
imag\_PS-Ymag & --- & --- & --- & --- & --- & 0.0043 & 0.0163 \\
imag\_PS-Hmag & --- & --- & --- & --- & --- & 0.0049 & 0.0042 \\
zmag\_PS-Kmag & --- & --- & --- & --- & --- & 0.0047 & 0.0148 \\
zmag\_PS-Jmag & --- & --- & --- & --- & --- & 0.0138 & 0.0185 \\
imag\_PS-Jmag & --- & --- & --- & --- & --- & 0.0024 & 0.0155 \\
Jmag-Kmag & --- & --- & --- & --- & --- & 0.0087 & 0.0204 \\
zmag\_PS-Ymag & --- & --- & --- & --- & --- & 0.0130 & 0.0146 \\
Ymag-Jmag & --- & --- & --- & --- & --- & 0.0103 & \textbf{0.0293} \\
gmag\_PS-Hmag & --- & --- & --- & --- & --- & 0.0030 & 0.0036 \\
gmag\_PS-Jmag & --- & --- & --- & --- & --- & 0.0040 & 0.0039 \\
Ymag-W2 & --- & --- & --- & --- & --- & 0.0130 & --- \\
gmag\_PS-Ymag & --- & --- & --- & --- & --- & 0.0012 & 0.0028 \\

rmag\_PS-Kmag & --- & --- & --- & --- & --- & 0.0042 & 0.0040 \\
gmag\_PS-Kmag & --- & --- & --- & --- & --- & 0.0020 & 0.0065 \\
rmag\_PS-Ymag & --- & --- & --- & --- & --- & 0.0032 & 0.0218 \\
rmag\_PS-Jmag & --- & --- & --- & --- & --- & 0.0058 & 0.0041 \\
rmag\_PS-Hmag & --- & --- & --- & --- & --- & 0.0032 & 0.0054 \\
Jmag-Hmag & --- & --- & --- & --- & --- & 0.0068 & \textbf{0.0272} \\
imag\_PS-Kmag & --- & --- & --- & --- & --- & 0.0037 & 0.0026 \\
Ymag-Kmag & --- & --- & --- & --- & --- & 0.0024 & 0.0116 \\
\end{longtable}

\twocolumn
\section{Uncertainty calibration}
\label{appendix:calibration}
\begin{figure}[H]
    \centering
    \includegraphics[width=0.7\linewidth]{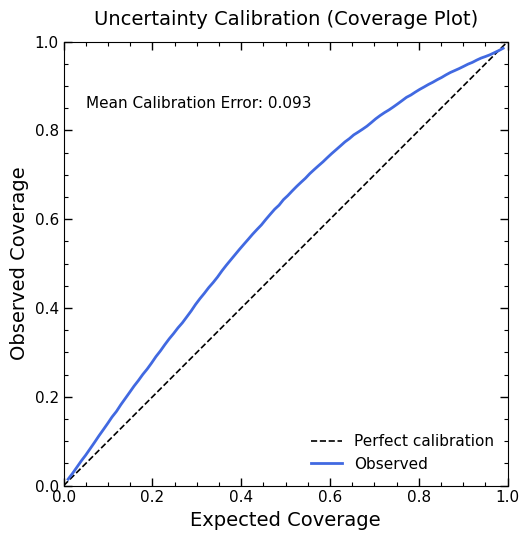}
    \caption{Uncertainty calibration (coverage plot) for the MDN with MC-dropout. The dashed black line indicates perfect calibration, while the solid blue line represents the observed coverage across different nominal confidence levels. The curve sits consistently above the diagonal, indicating that the predicted uncertainties are slightly conservative.}
    \label{fig:coverage_plot}
\end{figure}

Figure~\ref{fig:coverage_plot} shows the uncertainty calibration of the MDN with MC-dropout, evaluated via a coverage plot. For each expected coverage level $p$, we compute the empirical fraction of test targets falling within the central $p$-credible interval of the predictive distribution. The observed curve lies above the identity line across all confidence levels, with a mean calibration error of 0.093, indicating that the model is slightly underconfident: predictive intervals are conservative and slightly overestimate the true uncertainty. This behavior represents the partial redundancy between the aleatoric uncertainty captured by the mixture density output and the epistemic uncertainty estimated via MC-dropout. Since the intervals are on the side of over-coverage rather than under-coverage, the reported uncertainties can be regarded as reliable upper bounds on the prediction error. Although further recalibration techniques could be applied to tighten these uncertainty intervals, we choose to retain our slightly conservative estimates to ensure a more robust error bound for downstream catalog-level applications.

\section{DRW lightcurve simulations}
\label{appendix:simulations}
To understand the impact of applying single-band variability constraints on photo-$z$ estimation, we simulated DRW light curves for a subset of 8\,237 objects from our testing sample. These objects were selected based on the availability of black hole masses ($M_{\mathrm{BH}}$) from the SDSS \citep{Shen2011}.

The simulated light curves were generated following the methodology described in \citet{VARPZ}. For each object, we fitted the SED using the ``$grizYJHKW1W2$'' photometric data at their spectroscopic redshifts to estimate the absolute $i$-band magnitude ($M_{i}$) and the AGN flux fraction ($R_{\mathrm{AGN}}$). These derived values, along with the known $M_{\mathrm{BH}}$ measurements, were then applied to the DRW scaling relations prescribed in \citet{VARPZ} to calculate the characteristic damping timescale ($\tau$) and the asymptotic variability amplitude ($SF_{\infty}$). Using these parameters, we generated idealized ZTF-like \textit{g}-band DRW light curves that model pure continuum variability, without accounting for contributions from broad emission line region (BLR) reverberation or host-galaxy contamination. The simulated cadence and photometric uncertainties were drawn directly from the typical properties of actual ZTF \textit{g}-band observations.
\begin{figure}[H]
    \centering
    \includegraphics[width=1\linewidth]{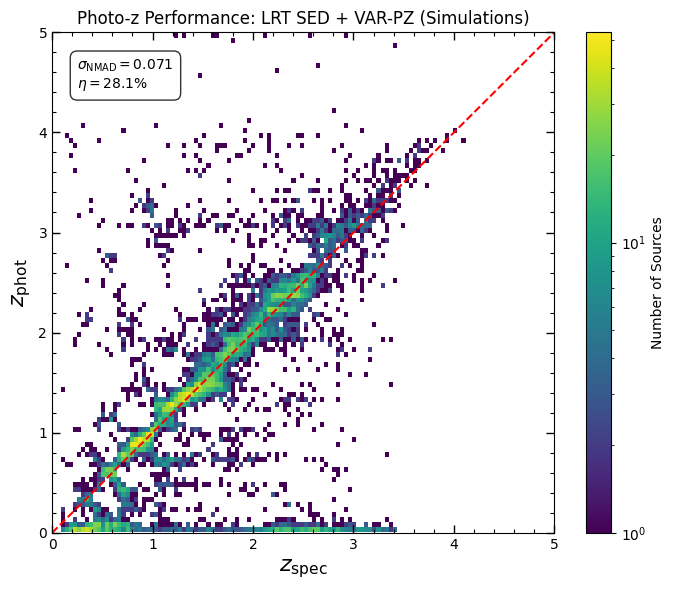}
   \caption{Performance of the \texttt{LRT}+\texttt{VAR-PZ} photo-$z$ framework when applied to simulated \textit{g}-band DRW light curves for 8237 objects from the testing sample ($\sigma_{\mathrm{NMAD}}=0.071$, $\eta=28.1\%$). The red dashed line indicates the one-to-one relation.}
    \label{fig:simulated_varpz}
\end{figure}
We applied the \texttt{LRT} SED template-fitting and the combined \texttt{LRT}+\texttt{VAR-PZ} framework to this simulated sample. For the \texttt{LRT}-only predictions, we obtain $\sigma_{\mathrm{NMAD}}=0.069$ and $\eta=27.6\%$. As shown in Figure~\ref{fig:simulated_varpz}, applying the combined \texttt{LRT}+\texttt{VAR-PZ} framework yields $\sigma_{\mathrm{NMAD}}=0.071$ and $\eta=28.1\%$.

The marginal worsening of $\sim0.5$ percentage points in the outlier fraction when \texttt{VAR-PZ} priors are applied, compared to the $\sim10.7$ percentage points degradation observed with real data (Section~\ref{sec:comparison}), can be explained by the idealized nature of the simulated light curves. Because the simulated data are generated from a pure DRW process with well-defined and consistent photometric uncertainties, the DRW model parameters are recovered more reliably, and the resulting variability priors introduce minimal bias into the photo-$z$ estimation. In contrast, real ZTF light curves are affected by observational systematics, non-DRW variability components, and heterogeneous photometric uncertainties. These observational realities can lead to degenerate DRW parameter solutions when only a single band is available, which in turn produce incorrect variability priors that actively degrade the SED-based photo-$z$ estimates.

\end{document}